\documentclass[aps,nofootinbib,amsmath,prd,showpacs,groupedaddress,12pt,notitlepage,superscriptaddress]{revtex4-1}

\usepackage{amsfonts}
\usepackage{amsmath}
\usepackage{hyperref}
\usepackage{epsfig}
\usepackage{latexsym}
\usepackage{paralist}
\usepackage{fancyhdr}
\usepackage{graphicx}

\numberwithin{equation}{section}

\usepackage[vcentermath]{youngtab}
\usepackage{young}
\usepackage{ytableau}
\usepackage{etex}
\usepackage{braket}
\usepackage{float}
\usepackage{slashed}
\usepackage{xcolor}

\usepackage{dcolumn} 

\makeatletter
\@addtoreset{equation}{section}

\makeatother

\def\bea{\begin{eqnarray}} 
\def\eea{\end{eqnarray}}
\def\be{\begin{equation}} 
\def\ee{\end{equation}} 
\def\ba{\begin{array}}
\def\ea{\end{array}} 
\def\nn{\nonumber}

\def\be{\begin{equation}}
\def\ee{\end{equation}}
\def\bea{\begin{eqnarray}}
\def\eea{\end{eqnarray}}

\def\nn{\nonumber}
\def\ep{\epsilon}

\renewcommand{\thefootnote}{\fnsymbol{footnote}}

\let\oldtitle\title
\renewcommand{\title}[1]{\oldtitle{\color{blue}{#1}}}

\let\oldeqref\eqref
\let\oldcite\cite
\renewcommand{\eqref}[1]{{\color{blue}\oldeqref{#1}}}
\renewcommand{\cite}[1]{{\color{blue}\oldcite{#1}}}

\makeatletter
\let\reftagform@=\tagform@
\def\tagform@#1{\maketag@@@{\ignorespaces\textcolor{blue}{(\ignorespaces #1 \unskip\@@italiccorr \ignorespaces)\ignorespaces}}}
\makeatother

\makeatletter
\renewcommand{\p@subsection}{}
\renewcommand{\p@subsubsection}{}
\makeatother

\usepackage{mathpazo}
\usepackage{amsmath}

\begin{document}

\title{Leading CFT constraints on multi-critical models in $d>2$}

\author{A.\ Codello}
\email{codello@cp3-origins.net}
\affiliation{CP$^3$-Origins, 
University of Southern Denmark,
Campusvej 55, 5230 Odense M, Denmark}
\affiliation{INFN - Sezione di Bologna, via Irnerio 46, 40126 Bologna, Italy}

\author{M.\ Safari}
\email{safari@bo.infn.it}
\affiliation{INFN - Sezione di Bologna, via Irnerio 46, 40126 Bologna, Italy}
\affiliation{
Dipartimento di Fisica e Astronomia,
via Irnerio 46, 40126 Bologna, Italy}

\author{G.\ P.\ Vacca}
\email{vacca@bo.infn.it}
\affiliation{INFN - Sezione di Bologna, via Irnerio 46, 40126 Bologna, Italy}

\author{O.\ Zanusso}
\email{omar.zanusso@uni-jena.de}
\affiliation{
Theoretisch-Physikalisches Institut, Friedrich-Schiller-Universit\"{a}t Jena,
Max-Wien-Platz 1, 07743 Jena, Germany}
\affiliation{INFN - Sezione di Bologna, via Irnerio 46, 40126 Bologna, Italy}

\begin{abstract}
\vspace{3mm}
We consider the family of renormalizable scalar QFTs with self-interacting potentials of highest monomial $\phi^{m}$ below their upper critical dimensions $d_c=\frac{2m}{m-2}$,
and study them using a combination of CFT constraints, Schwinger-Dyson equation and the free theory behavior at the upper critical dimension. 
For even integers $m \ge 4$ these theories coincide with the Landau-Ginzburg description of multi-critical phenomena and interpolate with the unitary minimal models in $d=2$,
while for odd $m$ the theories are non-unitary and start at $m=3$ with the {\tt Lee-Yang} universality class.
For all the even potentials and for the {\tt Lee-Yang} universality class, we show how the assumption of conformal invariance is enough to compute the scaling dimensions of the local operators $\phi^k$
and of some families of structure constants in either the coupling's or the $\epsilon$-expansion.
For all other odd potentials we express some scaling dimensions and structure constants in the coupling's expansion.
\end{abstract}

\pacs{}
\maketitle

\renewcommand{\thefootnote}{\arabic{footnote}}
\setcounter{footnote}{0}

\section{Introduction} \label{Section_introduction}

The past two years have seen the development of some new application of conformal field theory (CFT) methods
to the study of critical models in dimension bigger than two and, more specifically, close to their upper critical dimensions \cite{Rychkov:2015naa}.
The simple requirement that a theory is conformal invariant at a critical point, rather than simply scale invariant, strongly constrains the form of its correlators \cite{Nakayama:2013is}
and allows to write several nontrivial relations among them \cite{Dolan:2000ut}.
The two key ideas behind this approach are
to achieve consistency between conformal symmetry and the equations of motion through the use of the operatorial Schwinger-Dyson equations (SDE),
and to ensure regularity with the Gaussian theory when the dimension approaches its upper critical value in a limiting procedure.
Such a method has been able to reproduce the leading results for the $\epsilon$-expansion of the {\tt Ising}, {\tt Lee-Yang}, and {\tt Tricritical} Ising universality classes.
These results are very amusing in that none of the standard methods of quantum field theory (QFT) are used, including perturbation theory and the renormalization group, but just the knowledge of free (Gaussian) theory results for the correlators given by the Wick contractions.
These achievements thus point at the idea that CFT might work as a fully consistent replacement of the standard methods when critical properties are under investigation.

We will be interested in generalizing this idea to theories governed by the general $\phi^m$ potential.
In a Ginzburg-Landau description their action is
\begin{equation}\label{microscopic_action_GL}
\begin{split}
 S[\phi] & = \int {\rm d}^dx \Bigl\{\frac{1}{2}\partial_\mu\phi\partial^\mu\phi+\frac{g}{m!}\phi^{m} + \sum^{m-1}_{k=0}\frac{g_k}{k!} \phi^k\Bigr\}\,,
\end{split}
\end{equation}
for $m$ a natural number bigger than two. These models can be divided into two classes: On the one hand if $m=2n$, i.e.\ even, they are the so-called multi-critical models
which are protected by a $\mathbb{Z}_2$ parity ($\phi\to-\phi$) and include both the {\tt Ising} ($m=4$) and {\tt Tricritical} ($m=6$) universality classes as the first special cases.\footnote{
We follow the convention that universality classes such as {\tt Ising}'s are denoted with typeset font,
therefore the spin $\pm1$ Ising model at criticality is only one specific realization of the {\tt Ising} universality class
and the two should not generally be confused. The paper will deal with universality classes to a greater extent.
}
In the Landau-Ginzburg approach
the $\phi^{2n}$ effective potential describes a statistical system with a phase-transition that can be reached
by opportunely tuning the \emph{coupling} $g$ to a positive value, and in which $n$ distinct minima of the potential become degenerate \cite{Zamolodchikov:1987ti}.
On the other hand if $m=2n+1$, that is odd, \eqref{microscopic_action_GL} represents a sequence of multi-critical non-unitary theories
which are protected by a generalization of parity and include the {\tt Lee-Yang} universality class ($m=3$) as first example.
The non-unitary nature manifests itself in that the critical value of the coupling $g$ must be a purely imaginary number for the odd potentials.
We will see in more detail at the beginning of the next Section why, within a CFT approach, all the subleading couplings $g_k$ of \eqref{microscopic_action_GL}
do not play a significant role in tuning the action to criticality, therefore for the moment we shall simply ignore them.

The upper critical dimension of \eqref{microscopic_action_GL} is defined as the dimension $d$ at which the coupling $g$ is canonically dimensionless
\begin{equation}\label{critical_dimension}
\begin{split}
 d_m & = \frac{2m}{m-2}\,.
\end{split}
\end{equation}
A simple application of the Ginzburg criterion confirms that above the upper critical dimension the statistical fluctuations are weak
and the physics of \eqref{microscopic_action_GL} is Gaussian and controlled by mean-field critical exponents,
while below the upper critical dimension the fluctuations are strong enough to change the scaling properties 
and to provide the field $\phi$ with an anomalous dimension.
In the latter case a consistent expansion for the critical exponents can be achieved by studying the system slightly below the upper critical dimension
\begin{equation}\label{critical_dimension_epsilon}
\begin{split}
 d & = d_m-\epsilon\,,
\end{split}
\end{equation}
which for small $\epsilon$ tames the fluctuations and provides all the physically interesting critical quantities in the form of a Taylor series in $\epsilon$.\footnote{
In the non unitary models, e.g.\ {\tt Lee-Yang} universality class, the critical coupling and some structure constants are actually expressed as series of integer powers of $\epsilon^{1/2}$.}

The most important critical exponents of all the aforementioned special cases ({\tt Ising}, {\tt Tricritical} and {\tt Lee-Yang})
are known to high orders of the $\epsilon$-expansion \cite{Kleinert:2001ax,deAlcantaraBonfim:1980pe,deAlcantaraBonfim:1981sy,Gracey:2015tta,Macfarlane:1974vp}.
The leading and next-to-leading contributions in the $\epsilon$-expansion of \eqref{microscopic_action_GL} are known in general for all the even potentials $m=2n$
thanks to the application of standard perturbation theory, $\overline{\rm MS}$-methods and renormalization group analysis \cite{osborn_07}, while less is known for the odd potentials.
To underline how interesting and unexpected the results of \cite{osborn_07} for the even potentials are, let us point out that for $n\geq 3$ the leading contributions arise from \emph{multiloop computations},
and that for $n\geq 4$ the divergences are subtracted as poles of the \emph{fractional} dimensions $d_{2n}$ of \eqref{critical_dimension}!

Another interesting property is that the even models are known to interpolate in $d=2$ with the \emph{unitary} minimal CFTs ${\cal M}(p,p+1)$ for $p=1+m/2$,
which arise from the representations of the infinite dimensional Virasoro algebra \cite{Belavin:1984vu}.
Similarly, there are speculations \cite{Zambelli:2016cbw} pointing at the fact that the non-unitary models might interpolate with the sequence of minimal non-unitary multi-critical theories
${\cal M}(2,m+2)$ studied in \cite{Belavin:2003pu}. This is established for the {\tt Lee-Yang} case $m=3$ \cite{Cardy:1985yy}.
It is thus legitimate to generalize the arguments made for $m=3$ in \cite{Nii:2016lpa,Hasegawa:2016piv}, for $m=4$ in \cite{Rychkov:2015naa,Nakayama:2016cim,Nii:2016lpa} and for $m=6$ in \cite{Basu:2015gpa,Nii:2016lpa},
and assume that for each value of $m$ the multi-critical models at the critical point are conformal field theories for any dimension $2\leq d \leq d_{m}$.
The straightforward question that we will dare to answer in this paper is: how proficient will the Dyson-Schwinger consistency be in determining the critical properties of \eqref{microscopic_action_GL}?

The paper is organized as follows:
in Sect.\ \ref{section_SDE_and_CFT} we briefly summarize the main features of the Schwinger-Dyson consistency condition as well as some important property of CFT.
In Sect.s \ref{section-even-potentials} and \ref{section-odd-potentials} we treat the cases of even and odd potentials respectively.
All the results of these two Sections are summarized in the Subsections \ref{summary_even} and \ref{summary_odd}.
In Sect.\ \ref{conclusions} we attempt a unified conclusion and give some future prospects.
The Appendices collect some formulas which are very useful for our manipulations,
in particular Appendix \ref{free} deals extensively with the free theory in arbitrary dimension and the counting of the Wick theorem,
while Appendix \ref{lapla} collects few relations involving the action of the Laplacian on the CFT correlators.

\section{Schwinger-Dyson consistency and CFT}\label{section_SDE_and_CFT}

We dedicate this Section to a brief but more technical introduction to the application of the Schwinger-Dyson consistency condition in CFT.
Furthermore, some formulas of Sect.\ \ref{Section_introduction} necessitate further clarifications for their application to CFT, therefore there will be some slight overlapping with the previous Section.
Let us begin by introducing the action of the scalar $\phi^{m}$-theory
\begin{equation}\label{microscopic_action}
S[\phi] =\int d^d x \,  \Bigl\{ \frac{1}{2} (\partial \phi)^2 +\mu ^{\left(\frac{m}{2}-1\right) \epsilon}\frac{g }{m!} \phi^{m}   \Bigr\}\,,
\end{equation}
in $d$ dimensions, for $d$ sufficiently close to the upper critical dimension as in Eqs.\ \eqref{critical_dimension} and \eqref{critical_dimension_epsilon}.
The careful reader should have noticed several important details in comparing \eqref{microscopic_action} with \eqref{microscopic_action_GL}. In \eqref{microscopic_action} we introduced a reference (mass) scale $\mu$ which makes the almost marginal coupling $g$ dimensionless for any $d$.
The presence of the mass scale $\mu$ underlies the fact that the action \eqref{microscopic_action} is not conformal invariant for all values of $g$,
which in fact must be tuned to its critical value as will be done later in the paper.
Nevertheless, we could exclude all the strictly dimensionful couplings $g_k$ that appeared in \eqref{microscopic_action_GL} from \eqref{microscopic_action}.
The reason is that, since we are interested in the underlying conformal theory, which by definition does not depend on external scales, all couplings with positive mass dimension must vanish at criticality. This multi-critical tuning corresponds to the point in which, for example, all the $n$ different phases of a $\phi^{2n}$ theory coexist. 

Before diving more deeply into some technical details, it is worth noting that, with the exception of the cases $m=3,4$ and $6$, 
the upper critical dimension $d_m$ is a rational number.
More generally, after the displacement by $\epsilon$ all the theories will live in the arbitrarily real dimension $d=d_m-\epsilon$.
Theories living in continuous dimensions have already been investigated as CFT with conformal bootstrap methods \cite{El-Showk:2013nia}:
They are now believed to violate unitarity through the appearance of complex conjugate pairs of scaling dimensions,
which are probably related to ``evanescent'' operators that couple to the spectrum only at non-integer dimensionalities and are associated to states with negative norm \cite{Hogervorst:2015akt}.
While this is a very interesting line of research which deserves further investigation, we shall not deal with these aspects
and assume that conformal symmetry, unitary or not unitary, is realized for any value of the dimension $d$.\footnote{
Scale invariance seems to imply conformal invariance for several physically interesting critical models, especially in even dimensional cases.
There is also a pragmatic evidence, due to the results from conformal bootstrap program, that this is true for the $d=3$ {\tt Ising} universality class. 
This evidence has been recently supported at theoretical level~\cite{Delamotte:2015aaa}.}

The key idea of \cite{Rychkov:2015naa} is that all the CFT data of \eqref{microscopic_action} must interpolate with that of the Gaussian theory in the limit $\epsilon\to 0$.
We set some notation by defining the scaling dimensions for the field $\phi$ and the composite operators $\phi^m$ of an interacting scalar theory in $d$ dimensions.
Let the canonical dimension of $\phi$ be
\begin{align}
 \delta=\frac{d}{2}-1=\delta_m-\frac{\epsilon}{2}\,,
 \qquad {\rm with} \qquad \delta_m=\frac{2}{m-2}\,,
\end{align}
and the scaling dimensions of $\phi$ and $\phi^k$ be respectively
\begin{align}
 \Delta_1  \equiv \Delta_{\phi} =\delta+\gamma_1
 \qquad {\rm and} \qquad
 \Delta_k  \equiv \Delta_{\phi^k} =k \,\delta +\gamma_k\,.
\end{align}
The $\gamma$-terms represent the corrections from the canonical scaling dimensions $\delta$ and $k \,\delta$,
and therefore must be proportional to some power of $g$ or $\epsilon$ to ensure consistency of the Gaussian limit.

The Schwinger-Dyson equations (SDE) generalize the notion of equations of motion of \eqref{microscopic_action} at a functional and at an operatorial level.
Neglecting contact terms, any insertion of the equations of motion in a correlator constructed with a string of operators 
returns zero.
In practice, for any state of the CFT and for any list of operators $O_i$ the relation
\begin{align}
\left\langle\frac{\delta S}{\delta\phi}(x) \, O_1(y) \, O_2 (z) \dots\right\rangle  = 0
\label{prop}
\end{align}
holds. 
In general the SDE are constructed with renormalized quantities where explicit $\epsilon$-dependences do appear through the renormalized coupling in $S[\phi]$.
However, at the lowest order one can use the relation
\begin{align}\label{cft-sde}
\braket{\Box_x \phi(x)  O_1(y)  O_2 (z) \dots } =  \frac{g }{(m\!-\!1)!} \braket{\phi^{m-1}(x) O_1(y)  O_2 (z) \dots }
\end{align}
at tree level.
Thanks to the Schwinger-Dyson equation one can deduce that in the interacting CFT the operator $\phi$ and $\phi^k$
with $k\ne m\!-\!1$ are primaries, while the operator $\phi^{m-1}$ is a descendant.\footnote{
A descendant operator in $d > 2$ is the derivative of a primary operator, which is annihilated by the generator of the special conformal transformations.
We shall not be concerned with the higher complexity of the $d=2$ case.
}
In other words, the interacting CFT enjoys one less independent operator, that is $\phi^{m-1}$,
and a recombination of the conformal multiplets must take place.
In particular, the scaling dimensions of $\phi$ and $\phi^{m-1}$ must be constrained
\begin{equation}
\Delta_{m-1}=\Delta_1+2 \quad \Longrightarrow \quad  \gamma_{m-1}=\gamma_1+(m\!-\!2) \frac{\epsilon}{2}\,.
\label{special_anomalous}
\end{equation}

Furthermore, conformal symmetry greatly constrains the correlators appearing on both sides of the SDE.
It is possible to find a basis $O_a$ of scalar primary operators with scaling dimensions $\Delta_a$ whose two point correlators are diagonal
\begin{align}\label{cft-2pf}
\braket{O_a(x)  O_b(y)} =\frac{c_a \, \delta_{ab}} {|x-y|^{2 \Delta_a}} \,,
\end{align}
(no summation over $a$)
where we denoted as $c_a$ the general non-negative normalization factors which can in principle be set to one.
However, for the moment, we will find it more convenient to work with the natural normalization of the Gaussian theory, that is induced by Wick counting.
The tree-point correlator for scalar primary operators is even more constrained by conformal symmetry and reads
\begin{align}\label{cft-3pf}
\braket{O_a(x)  O_b(y) O_c(z)} =\frac{C_{abc} }{|x-y|^{\Delta_{a}+\Delta_{b}-\Delta_{c}} 
|y-z|^{\Delta_{b}+\Delta_{c}-\Delta_{a}} |z-x|^{\Delta_{c}+\Delta_{a}-\Delta_{b}}} \,.
\end{align}
where $C_{abc}=C_{a,b,c} $ are known as the structure constants of the CFT (we will adopt the notation with the commas whenever a potential notational ambiguity arises).
Our CFTs are completely and uniquely specified by providing the scaling dimensions $\Delta_a$ and the structure constants $C_{abc}$,
which together are known as \emph{CFT data} and which are for obvious reasons paramount target of any computation.

Our goal is to extract the leading informations for a part of the conformal data of all the multi-critical CFT (including scaling dimensions and structure constants). Our results can be seen as the first step before investigating such an infinite family of multi-critical theories at an interacting fixed point in $d_m\!-\!\ep$ dimensions as a power series in $\ep$, eventually with also conformal bootstrap techniques. 

\section{\boldmath{$\phi^{2n}$}-theory in \boldmath{$d=d_{2n}-\epsilon$} dimensions} 
\label{section-even-potentials}

This Section is dedicated to the investigation of the even potentials $\phi^{2n}$ in $d=d_{2n}-\epsilon$ dimensions
which arise as the special case $m=2n$ of \eqref{microscopic_action} and which are explicitly $\mathbb{Z}_2$ symmetric under parity.
Throughout this Section we will reserve the symbol $n$ exclusively for the natural number bigger than one,
which is half of $m$ whenever $m$ is even.
Naturally, $n$ is in one-to-one correspondence with the model and labels its criticality, which is the number of degenerate ground states at the critical point.
To give the results some context, we find useful to explicitly list the first few critical dimensions:
Starting from the case $n=2$ that corresponds to the {\tt Ising} universality class, the upper critical dimensions are
\begin{align}
d_{2n} &= \frac{2n}{n-1}= 4 \,,\, 3 \,,\, \frac{8}{3} \,,\, \frac{5}{2} \,,\, \frac{12}{5} \,,\, \dots \,,\, 2\,.
\end{align}
They become purely rational numbers starting from $d_8=\frac{8}{3}$,  corresponding to the {\tt Tetra\-cri\-ti\-cal} universality class, which in $d=2$ describes  the 3-states Potts model at criticality. 
In the limit $n\to\infty$ the critical dimensions tend to two, that is the dimensionality for which the canonical dimension of the field is zero and all couplings are canonically marginal.
From our point of view, the study of the even models is particularly interesting because it allows for a direct and very general comparison of our results with those obtained in \cite{osborn_07},
and serves as a testing ground for the entire method.

In the first part of this Section, we will kickstart the computation by obtaining the anomalous dimension for the field $\phi$ by using a constraint
which comes from the consistency of the two point function \eqref{cft-2pf} with the SDE \eqref{cft-sde} in the limit $\epsilon\to0$.
Then we will repeat the process by requiring consistency of the three point function \eqref{cft-3pf} to determine the scaling dimensions of all the composite operators $\phi^k$.
We will see that $\gamma_2$, which is related to the anomalous scaling of the correlation length, requires a separate discussion for all the theories with $n>2$.
In the second part of this Section we will determine several of the structure constants $C_{abc}$ which appeared in \eqref{cft-3pf}.
In particular, we will mostly concentrate on those that are not present at zeroth order in $\epsilon$
and are thus generated at quantum level.
In the third part we will exploit the fact that the scaling dimension of the $\phi^{2n-1}$ descendant operator can be computed in two different ways
and use it to find a critical value for the coupling $g$ as a function of $\ep$. We will also manifest some explicit relation with the standard perturbation theory of \cite{osborn_07}.
All the results are summarized at the end of the Section.

\subsection{Anomalous dimensions }

Our first goal is the computation of the leading order (LO) anomalous dimensions of the field $\gamma_1$ and of the composite operators $\gamma_k$ with $k\geq2$.
By LO we will generally mean leading order in $g$ and in $\epsilon$. Only when an explicit relation $g(\epsilon)$ will be available (as for even potentials in Section~\ref{gofederiv}), leading order will mean  leading order in $\epsilon$.  

We  start with a simple analysis of the two point function 
that will directly uncover a precise leading order relation between $\gamma_1$ and the coupling $g$.
The determination of $\gamma_2$ requires the analysis of 
three point function $\braket{ \phi \,\phi\,\phi^2}$ and is a bit more involved for $n>2$. 
Finally we shall be able to obtain the anomalous dimensions $\gamma_k$ with $k\geq n$ from the study of $\braket{ \phi \,\phi^k \phi^{k+1}}$.
In these first computations we will proceed step by step in order to explain the details of the method we employ. We assume the knowledge of the free theory correlators as detailed in Appendix~\ref{free}.

\subsubsection{Warm-up: $\gamma_1$}

Let us consider in $d$ dimensions the propagator of the interacting theory
\begin{align}
\braket{\phi(x)  \phi(y)} = \frac{C}{|x-y|^{2 \Delta_1}} \,.
\label{2pfdress}
\end{align}
The renormalized result for the CFT (e.g.~at the fixed point) is characterized by a normalization which at lowest order is given by the free theory one $C=c+O(g)$, where $c$ is given in Eq.~\eqref{normprop}. Thus we will make the replacement $C\to c$ everywhere from now on.

On applying first the SDE in one point one shows that $\gamma_1$ is at least of order $g^2$.
Then applying the SDE also to the second point gives the 
leading expression for $\gamma_1$ in terms of $g$.
Acting with a Laplacian in Eq.~\eqref{2pfdress} using
${\Box_x |x\!-\!y|^{-2 \Delta_1}=2 \Delta_1 (2 \Delta_1\!-\!2\delta) |x\!-\!y|^{-2 \Delta_1-2}}$
and recalling $\Delta_1=\delta+\gamma_1$ gives 
\begin{equation}
 \Box_x \braket{ \phi(x)  \phi(y)} = \Box_x \frac{c}{ |x-y|^{2 \Delta_1}}
=c\,\frac{ 4(\delta + \gamma_1)\gamma_1}{|x-y|^{2(1+\delta+\gamma_1)}}
\overset{{\rm LO}}{=} {\frac{4}{n-1}} \gamma_1 \frac{c}{|x-y|^{2+{\frac{2}{n-1}}}}\,.
\end{equation}
In this case the determination of the leading order contribution amounted to the substitutions $4(\delta + \gamma_1)\gamma_1 \to 4\delta_{2n}  \gamma_1$  in the numerator and $1+\delta+\gamma_1 \to 1+\delta_{2n}$ in the  denominator, where $\delta_{2n} = \frac{1}{n-1}$ is the upper critical dimension value of $\delta$.
Computing instead the above expression using the SDE one finds
\begin{align}
\braket{ \Box_x \phi(x) \phi(y)} &= \frac{g \mu^{(n\!-\!1)\epsilon}}{(2n\!-\!1)!} \ \braket{\phi^{2n-1}(x)  \phi(y)} = O(g^2)\,.
\end{align}
This is because the two point function on the right hand side vanishes in the free theory.
Therefore $\gamma_1$ is at least of order $g^2$.
To obtain another useful relation one acts with a Laplacian in $y$, computes explicitly the expression in terms of the anomalous dimensions and compares the result to the one obtained applying the SDE.
From the first computation one gets
\begin{align}
 \Box_x \Box_y \braket{\phi(x)  \phi(y)} &= \Box_x \Box_y \frac{c}{|x-y|^{2 \Delta_1}}
= c\,\frac{2\Delta_1(2\Delta_1 \!+\!2) (2\Delta_1\! -\!2\delta)  (2\Delta_1\!+\!2\!-\!2\delta)}{|x-y|^{2\Delta_1+4}} \nonumber  \\
&\overset{{\rm LO}}{=} \frac{16 n}{(n\!-\!1)^2} \gamma_1 \frac{c }{ |x-y|^{4+\frac{2}{n-1}}} \,,
\label{2pfbb}
\end{align}
where to determine the LO contributions we used $2\Delta_1(2\Delta_1 \!+\!2) (2\Delta_1\! -\!2\delta)  (2\Delta_1\!+\!2\!-\!2\delta) = 16 (\delta +\gamma_1)(\delta+1+\gamma_1)\gamma_1(1+\gamma_1)  \to 16 \delta_{2n}(\delta_{2n}+1)\gamma_1 $ in the numerator and $2\Delta_1+4 \to 2\delta_{2n}+4$ in the exponent in the denominator.
Applying the SDE and using  the free result for the two point function of Eq.~\eqref{2pf} of Appendix~\ref{free} gives instead
\begin{align}
\braket{ \Box_x \phi(x) \Box_y \phi(y)} &= \left( \frac{g \mu^{(n-1)\epsilon}}{(2n\!-\!1)!} \right)^2 \braket{\phi^{2n-1}(x)  \phi^{2n-1}(y)} 
\overset{{\rm LO}}{=}\frac{g^2}{(2n\!-\!1)!} \frac{c^{2n-1}}{|x-y|^{4+\frac{2}{n-1}}} \,.
\label{2pfSDE}
\end{align}
By comparing Eq.~\eqref{2pfbb} and Eq.~\eqref{2pfSDE}  one immediately finds the leading contribution to the anomalous dimension
\begin{align}
\gamma_1 
=c^{2(n-1)} \frac{(n\!-\!1)^2}{8 (2n)!} g^2+O(g^3) \,.
\end{align}
Using the fact that
\be
c = \frac{\Gamma(\delta_{2n})}{4\pi^{1+\delta_{2n}} }
\ee 
we find the explicit formula
\begin{align}
\gamma_1 
=  \frac{2  (n\!-\!1)^2}{ (2n)!} \Gamma \Big(\!{\textstyle \frac{1}{n-1}}\!\Big)^{2(n-1)} \frac{g^2}{(4\pi)^{2n}}+O(g^3)\,,
\label{gamma1}
\end{align}
which agrees with the perturbative result~\cite{osborn_07}.

\subsubsection{Climbing up: $\gamma_2$}
To determine $\gamma_2$ we need to consider the three point functions. The simplest correlator where it appears is
\begin{align}
\braket{\phi(x) \phi (y) \phi^{2} (z)  } &= \frac{ C_{112}}{ |x-y|^{2\Delta_{1}\!-\!\Delta_{2}} |y-z|^{\Delta_{2}} |z-x|^{\Delta_{2}}}\,.
\label{3pf11k}
\end{align}
In this correlator the SDE can be used twice at the points $x$ and $y$.
The action of one Laplacian can be easily obtained from Eq.~\eqref{box3pf1} given in Appendix \ref{lapla}
by setting  $\alpha_1= 2\Delta_{1}\!-\!\Delta_{2} = 2\gamma_1\!-\!\gamma_2$ and $\alpha_2=\alpha_3 = \Delta_{2}=2\delta\!+\!\gamma_2$
\begin{align*}
 &\Box_x \frac{1}{|x-y|^{2\Delta_{1}-\Delta_{2}} |y-z|^{\Delta_{2}} |z-x|^{\Delta_{2}}}
 = \frac{ 2(2\gamma_1\!-\!\gamma_2)\gamma_1 }{|x-y|^{2+2\gamma_1-\gamma_2} |y-z|^{2\delta+\gamma_2} |x-z|^{2\delta+\gamma_2}} \nonumber \\
 &+  \frac{2(2\delta\!+\!\gamma_2)\gamma_1}{|x-y|^{2\gamma_1-\gamma_2} |y-z|^{2\delta+\gamma_2} |x-z|^{2\delta+\gamma_2+2}}
 -  \frac{(2\gamma_1\!-\!\gamma_2)(2\delta\!+\!\gamma_2)}{|x-y|^{2+2\gamma_1\!-\!\gamma_2} |y-z|^{2\delta+\gamma_2-2} |x-z|^{2\delta+\gamma_2+2}}\,.
\end{align*}
From this expression we easily determine the leading order contributions
\begin{align}
\Box_x \braket{ \phi(x) \phi (y) \phi^{2} (z)  }&\overset{{\rm LO}}{=}   \frac{{\frac{8c^2}{n-1}} \gamma_1}{|y-z|^{{\frac{2}{n-1}} } |z-x|^{\frac{2 n}{n-1}}} - \frac{{\frac{4c^2}{n-1}}(2\gamma_1\!-\!\gamma_2) }{|x-y|^{2 } |y-z|^{\color{black}2 {\frac{2-n}{n-1}}} |z-x|^{\frac{2 n}{n-1}}} \,,
\label{box3pf1gamma2}
\end{align}
where we also made the leading order substitution $C_{112} \to C_{112}^{\mathrm{free}} = 2 c ^2$.
This expression should match the one obtained by applying the SDE
\begin{align}
 \braket{\Box_x \phi(x) \phi (y) \phi^{2} (z)  } = \frac{g \mu^{(n\!-\!1)\epsilon}}{(2n\!-\!1)!} \braket{\phi^{2n-1}(x) \phi(y)  \phi^2(z) }\overset{{\rm LO}}{=} g\, \frac{  \delta_{n,2} \,c^3}{|x-y|^{2} |z-x|^{4}}\,,
\end{align}
where we used $C_{312}^{\rm free} = 6 c^3$. 
Therefore, comparison with Eq.\ \eqref{box3pf1gamma2} shows that $\gamma_2 \sim O(g^2)$ for $n>2$ while it is of order $O(g)$ only for $n=2$, for which case it is determined by the following expression
\begin{equation}
\gamma_2 = \frac{g}{(4\pi)^2} + O(g^2) \,,\qquad n=2 \,.
\label{gamma2n2}
\end{equation}
In order to find the leading value of $\gamma_2$ in the general case $n>2$ we act with the second Laplacian in $y$.
Using Eq.\ \eqref{boxbox3pf11} from the Appendix \ref{lapla}
and keeping the leading contributions one finds (we skip the intermediate steps)
\begin{align}
\Box_x \Box_y \braket{\phi(x) \phi(y)  \phi^2(z) } &\overset{{\rm LO}}{=}
\frac{16(n\!-\!2)c^2}{(n\!-\!1)^2}\frac{ \gamma_2 \!-\! 2 \gamma_1 }{|x-y|^{4} |y-z|^{\frac{2}{n-1}} |z-x|^{\frac{2}{n-1}}}\,,
\end{align}
which we should compare with the leading order result obtained applying the SDE,
\begin{align}
\!\!\!\left( \frac{g \mu^{(n\!-\!1)\epsilon}}{(2n\!-\!1)!}\right)^2 \!\!\! \braket{\phi^{2n-1}(x) \phi^{2n-1}(y)  \phi^2(z) } &\overset{{\rm LO}}{=}
\frac{g^2 }{(2n\!-\!1)!^2} \frac{C_{2n-1,2n-1,2}^{\mathrm{free}}}{  |x-y|^{4} |y-z|^{\frac{2}{n-1}} |z-x|^{\frac{2}{n-1}}} \,,
\end{align}
so that by comparison we obtain
\begin{equation}
\gamma_2- 2\gamma_1 =\frac{ (n\!-\!1)^2}{16(n\!-\!2)(2n\!-\!1)!^2} \frac{C_{2n-1,2n-1,2}^{\mathrm{free}}}{c^2  } g^2+O(g^3)\,.
\end{equation}
Using the explicit expression for $\gamma_1$ given in Eq.~\eqref{gamma1} we find
\begin{align}
\gamma_2 = 
8  \frac{(n\!+\!1)(n\!-\!1)^3} {(n\!-\!2)(2n)!} \Gamma \left({\textstyle \frac{1}{n\!-\!1}}\right)^{\!2(n-1)} \frac{g^2}{(4\pi)^{2n}}+O(g^3)\,, \qquad n>2
\,.
\label{gamma2}
\end{align}
This quantity has not been reported in the perturbative results given in~\cite{osborn_07}.

\subsubsection{The general case: $\gamma_k$}

To determine $\gamma_k$ at first we could think to consider $\braket{\phi\, \phi \, \phi^{k} }$, but this correlator is zero in the free theory whenever $k>2$. To investigate all $k\geq 2$ we instead consider the following three point function
\begin{equation}
\braket{\phi(x) \phi^k (y) \phi^{k+1} (z)  } = \frac{C_{1,k,k+1}}{ |x-y|^{\Delta_{1}\!+\!\Delta_{k}\!-\!\Delta_{k+1}} |y-z|^{\Delta_{k}\!+\!\Delta_{k+1}\!-\!\Delta_{1}} |z-x|^{\Delta_{1}\!+\!\Delta_{k+1}\!-\!\Delta_{k}}}\,.
\label{3pf1kkp1}
\end{equation}
The general expression on the right hand side is valid for primary operators, that is for $k \ne 2n-2,2n-1$. Indeed for $k=2n-2, 2n-1$ other terms are present. Nevertheless, if one restrict the analysis to the lowest order, these extra terms which are subleading can be neglected and \eqref{3pf1kkp1} can be used also for these two cases, as will be discussed in the Subsection~\ref{gofederiv}.

The leading value for the normalization is obtained from the free theory approximation from the general
expression~\eqref{c3_free} and reads
\begin{align}
C_{1,k,k+1}^{\mathrm{free}} &= (k\!+\!1)! \, c^{k+1} \,.
\label{c1kkp1}
\end{align}
The main recursion relation can then be derived for $k\ge n\!-\!1$ applying a Laplacian in $x$ and exploiting the relation given by the SDE.
Using the relation \eqref{box3pf1} in Appendix ~\ref{lapla}
one can compute the action of a Laplacian in $x$ on the correlator \eqref{3pf1kkp1} for which, following the same reasoning of the previous Subsections, we find the following LO expression
\begin{align}
 \Box_x \braket{ \phi(x) \phi^k (y) \phi^{k+1} (z)  } &\overset{{\rm  LO}}{=} \frac{4 \gamma_1}{n\!-\!1} \frac{C_{1,k,k+1}}{|y-z|^{\frac{2k}{n-1}} |z-x|^{\frac{2n}{n-1}}} \nonumber \\
& + \,\frac{2}{n\!-\!1}(\gamma_{k+1}\!-\!\gamma_{k}\!-\!\gamma_1) \frac{C_{1,k,k+1}}{ |x-y|^{2} |y-z|^{\frac{2k}{n-1}-2} |z-x|^{\frac{2n}{n-1}}}\,.
\label{box1kkp1}
\end{align}
On the other hand using the SDE one gets
\begin{align}
\braket{ \Box_x \phi(x) \phi^k (y) \phi^{k+1} (z)  } &=\frac{g \mu^{(n-1)\epsilon}}{(2n\!-\!1)!} \braket{\phi^{2n-1}(x) \phi^k (y) \phi^{k+1} (z)  }\nonumber\\
& \overset{{\rm LO}}{=} \frac{g }{(2n\!-\!1)!} \frac{C_{2n-1,k,k+1}^{\mathrm{free}}}{ |x-y|^{2} |y-z|^{\frac{2k}{n-1}-2} |z-x|^{\frac{2n}{n-1}}}\,,
\label{sde1kkp1}
\end{align}
where
\begin{align}
C_{2n-1,k,k+1}^{{\rm free}} &= \frac{k! (k\!+\!1)! (2n\!-\!1)!}{(k\!-\!n\!+\!1)! (n-\!1\!)! \,n!} \,c^{k+n}\,, \quad  \quad k\ge n\!-\!1\,.
\end{align}
Vice versa when $k \le n\!-\!2$ the free correlator is zero and the the full correlator in Eq.~\eqref{sde1kkp1} is at least of order $O(g^2)$. 
The expression obtained from the SDE in Eq.~\eqref{sde1kkp1} has a leading term $O(g)$, and recalling from Eq.~\eqref{gamma1} that $\gamma_1= O(g^2)$,
one is forced to conclude that the first term in Eq.~\eqref{box1kkp1} is negligible and that $\gamma_{k+1} -\gamma_{k} = O(g)$. Then by comparing Eqs.~\eqref{box1kkp1} and~\eqref{sde1kkp1} one finds the recurrence relation
\begin{align}
\gamma_{k+1} -\gamma_{k}=\frac{2}{ (n\!-\!2)! \, n!} \frac{k!}{(k\!-\!n\!+\!1)!}
 \Gamma \left({\textstyle \frac{1}{n-1}}\right)^{(n-1)}\frac{g  }{(4 \pi)^n } +O(g^2)\,, \qquad k\ge n\!-\!1\,.
\end{align}
The recurrence relation for the anomalous dimensions associated to a difference of order $O(g)$ ceases to exists for $k\le n\!-\!2$ and is substituted by some relation involving $O(g^2)$ corrections.  Therefore we expect $\gamma_k = O(g^2)$ for $k\le n\!-\!1$. 
With this condition one can solve the recurrence relation to obtain
\begin{align}
\gamma_{k}=\frac{2 (n\!-\!1) }{ n!^2} \frac{k!}{(k\!-\!n)!}  \Gamma \left({\textstyle{\frac{1}{n-1}}}\right)^{n-1}\frac{ g }{(4 \pi)^n } +O(g^2)\,,   \qquad  k \ge n\,,
\label{gammakg}
\end{align}
which is in perfect agreement with the perturbative result~\cite{osborn_07}.
Note that in the case $n=2$ we correctly reproduce Eq.~\eqref{gamma2n2}. 

The above relation says that for $k\ge n-1$ the leading contribution to the anomalous dimensions is of $O(\epsilon)$. It is evident from our derivation of Eq.~\eqref{gammakg}, which is simply based on CFT invariance and the SDE, that this equation is valid for any $k$.  
In fact at this order, i.e.\ $O(\epsilon)$, one can also see from the point of view of  perturbative renormalization group that the anomalous dimensions \eqref{gammakg} are not affected by the mixing with derivative operators~\cite{CSVZ1}. The contribution from mixing with derivative operators may start only at next to leading order $O(\epsilon^2)$, and for $k \ge 2n$.

%



\subsection{Structure constants}

Besides the scaling dimensions, a CFT is also characterized by the structure constants of the three point correlators, which are related to the OPE coefficients.
We explore here the possibility to extract in the most generality some of them at leading order for the whole family of even universality classes.

In order to get some information from the three point functions using the Schwinger-Dyson equations we need to have one of the fields to appear with power one. The $\langle \phi\, \phi^k\phi^{k+1}\rangle$ are already explored and give information on the scaling dimensions $\Delta_i$. In the following we therefore concentrate on the rest of these correlation functions.

\subsubsection{Structure constants $C_{1,2k,2l-1}$} \label{C_{1,2k,2l-1}}
The remaining correlation functions consist of $\langle \phi \,\phi^k\phi^l\rangle$, $|k-l|\neq 1$. These vanish in the free theory, so they can give information on the structure constants $C_{1kl}$ and imply that these are at least proportional to the coupling or smaller. Now if $\langle \phi^{2n-1} \phi^k\phi^l\rangle$ also vanishes in the free theory it implies that $C_{1kl}$ are at least of order $O(g^2)$ and to find their value at leading order we need to know $\langle \phi^{2n-1} \phi^k\phi^l\rangle$ beyond free theory. Therefore we will not be able to extract the leading order information on $C_{1kl}$ this way, but for the case discussed in the next Subsection.

For $\langle \phi^{2n-1} \,\phi^k\phi^l\rangle$ not to vanish in the free theory we must have the following conditions. Since $2n-1$ is odd, either $k$ or $l$ must be even while the other must be odd, so we restrict ourselves to $\langle \phi^{2n-1}\, \phi^{2k}\,\phi^{2l-1}\rangle$, with $k,l\geq 1, n>1$. As previously discussed, the condition for this to be nonzero is 
\be 
\ba{l} k+l-n\geq 0 \\ l+n-k\geq 1 \\ k+n-l \geq 0\,. \ea
\ee
These are equivalent to $k+l\geq n, \, -n\leq k-l\leq n\!-\!1$. Furthermore we must have $l\neq k, k+1$ otherwise we will be back to the case $\langle \phi\, \phi^k\phi^{k+1}\rangle$ which is already studied. In summary, for $k,l$ satisfying the conditions
\be 
k+l\geq n, \quad 1\!-\!n\!\leq l-k\leq n, \quad l-k\neq 0 {\rm \,\,\, or \,\,\,} 1\,,
\label{opeset}
\ee
we can find the leading order ($O(g)$) structure constants $C_{1,2k,2l-1}$. One can use the SDE to write 
{\setlength\arraycolsep{2pt}
\bea 
&&  \langle \square_x\phi(x) \,\phi^{2k}(y)\; \phi^{2l-1}(z)  \rangle = \frac{g}{(2n\!-\!1)!} \langle \phi^{2n-1}(x) \,\phi^{2k}(y)\; \phi^{2l-1}(z)  \rangle   \\
&\stackrel{\mathrm{LO}}{=}& \frac{g}{(2n\!-\!1)!} \frac{C^{\mathrm{free}}_{2n-1,2k,2l-1}}{|x-y|^{\Delta_{2n-1}+\Delta_{2k}-\Delta_{2l-1}}|x-z|^{\Delta_{2n-1}+\Delta_{2l-1}-\Delta_{2k}}|y-z|^{\Delta_{2k}+\Delta_{2l-1}-\Delta_{2n-1}}}\,, \nn
\eea}%
which has been evaluated in the second line at leading order. On the other hand, applying the $\square_x$ to the correlation function $\langle \phi(x) \,\phi^{2k}(y)\;\phi^{2l-1}(z)\rangle$ one finds
{\setlength\arraycolsep{2pt}
\bea 
&& \hspace{-8mm} \square_x \langle \phi(x) \,\phi^{2k}(y)\; \phi^{2l-1}(z)  \rangle  \nn\\
&& = C_{1,2k,2l-1}\square_x \frac{1}{|x-y|^{\Delta_{1}+\Delta_{2k}-\Delta_{2l-1}}|x-z|^{\Delta_{1}+\Delta_{2l-1}-\Delta_{2k}}|y-z|^{\Delta_{2k}+\Delta_{2l-1}-\Delta_{1}}} \nn \\
&& \stackrel{\mathrm{LO}}{=} C_{1,2k,2l-1} \frac{(k-l)(k-l+1)(d_{2n}\!-\!2)^2}{|x-y|^{\Delta_{1}+\Delta_{2k}-\Delta_{2l-1}+2}|x-z|^{\Delta_{1}+\Delta_{2l-1}-\Delta_{2k}+2}|y-z|^{\Delta_{2k}+\Delta_{2l-1}-\Delta_{1}-2}}\,,
\eea}%
where the operator dimensions in the third line are understood as their leading order values. One readily sees, using the relation $\Delta_{2n-1} = \Delta_1+2$, that the denominators in the above two expressions are equal. Comparing the coefficients we find
\be 
C_{1,2k,2l-1} \stackrel{\mathrm{LO}}{=} \frac{g}{(2n-1)!} \frac{(n-1)^2C^{\mathrm{free}}_{2n-1,2k,2l-1} }{4(k-l)(k-l+1)}\,,
\label{c12k2lm1}
\ee
where
\be 
C^{\mathrm{free}}_{2n-1,2k,2l-1}  = \frac{(2n-1)!(2l-1)!(2k)!}{(n\!+\!l\!-\!k\!-\!1)!(k\!+\!n\!-\!l)!(k\!+\!l\!-\!n)!} \,c^{n+k+l-1}
\ee
and $c$ is the normalization of the free propagator given in Eq.~\eqref{normprop}.

\subsubsection{Structure constants $C_{1,1,2k}$}

The previous relation~\eqref{c12k2lm1} for $l=1$ gives two possible coefficients $C_{1,1,2k}\sim O(g)$ for $k=n\!-\!1,n$. We shall show in the following that one can find the leading behaviour of the other 
coefficients of the form $C_{1,1,2k}$ with $k$ in the range $2\leq k \leq 2n-1$, which turn out to be of order $O(g^2)$. These can be extracted from the analysis of the family of correlators considered in the previous Subsection
\begin{align}
\braket{\phi(x) \phi (y) \phi^{2k} (z)  } &=  \frac{C_{1,1,2k}}{|x-y|^{2\Delta_{1}\!-\!\Delta_{2k}} |y-z|^{\Delta_{2k}} |x-z|^{\Delta_{2k}}}\,, 
\label{3pf112k}
\end{align}
where $k>1$. Clearly the coefficients $C_{1,1,2k}$ for $k>1$ vanish in the free theory.
We proceed as before by acting on the above correlation function with two Laplacian operators in $x$ and $y$ and exploiting the SDE. Using the Eq.~\eqref{boxbox3pf11} of the Appendix we find at leading order
\begin{align}
\Box_x \Box_y \braket{\phi(x) \phi(y)  \phi^{2k}(z)} &\stackrel{\mathrm{LO}}{=} \frac{16k(k\!-\!1)(k\!-\!n)(k\!-\!n\!+\!1)}{(n\!-\!1)^4} \frac{C_{1,1,2k}}{|x\!-\!y|^{2(1-k)\delta_{2n}+4} |y\!-\!z|^{2k\delta_{2n}} |x\!-\!z|^{2k\delta_{2n}}}\,.
\end{align}
One can notice from this expression that the r.h.s.\ vanishes for $k = 1,n\!-\!1, n$. This means that for these values of $k$ the leading order expression will involve the anomalous dimensions and the present analysis will give relations involving these quantities. Restricting to the case $k\ne 1,n\!-\!1, n$, we compare the above equation  with the one obtained from applying the SDE
\begin{eqnarray}
\braket{\Box_x\phi(x)\Box_y\phi(y)\phi^{2k}(z)} &=& \frac{g^2 \mu^{2(n\!-\!1)\epsilon}}{(2n\!-\!1)!^2} \braket{\phi^{2n-1}(x) \phi^{2n-1}(y) \phi^{2k}(z)} \nn\\
&\stackrel{\mathrm{LO}}{=}& \frac{g^2 }{(2n\!-\!1)!^2} \frac{C_{2n-1,2n-1,2k}^{\mathrm{free}}}{|x-y|^{2(1-k)\delta_{2n}+4} |y-z|^{2k\delta_{2n}} |x-z|^{2k\delta_{2n}}}\,, \nn\\
\end{eqnarray}
so that we obtain
\be 
C_{1,1,2k} \stackrel{\mathrm{LO}}{=}\frac{g^2}{(2n-1)!^2} \frac{(n\!-\!1)^4C^{\mathrm{free}}_{2n-1,2n-1,2k}}{16k(k\!-\!1)(k\!-\!n)(k\!-\!n\!+\!1)}\,.
\ee
The structure constant on the r.h.s evaluated in the free theory is nonzero for $k\leq2n\!-\!1$ 
\be 
C^{\mathrm{free}}_{2n-1,2n-1,2k} = \frac{(2k)!(2n\!-\!1)!^2}{k!^2(2n\!-\!k\!-\!1)!} \,c^{2n+k-1}\,.
\ee
This gives
\be 
C_{1,1,2k} \stackrel{\mathrm{LO}}{=} \frac{(2k)! (n\!-\!1)^4\, c^{2n+k-1}}{16k(k\!-\!1)(k\!-\!n)(k\!-\!n\!+\!1)k!^2(2n\!-\!k\!-\!1)!} \,g^2 \,.
\label{c112k}
\ee
For higher values of $k$ one needs to know the correlation function $\braket{\phi^{2n-1} \phi^{2n-1} \phi^{2k}}$ beyond free theory, therefore it is not possible to extract the leading order $C_{1,1,2k}$ in this way. The range of validity for this formula is therefore $2\leq k\le 2n-1$ and $k\neq n-1,n$. As mentioned before, $1,n-1,n$ were excluded from the possible values $k$ can take in this Subsection, and will give information on the anomalous dimensions. The case $k=1$ has already been analysed in previous Subsections and gives $\gamma_2$. The other two cases $k=n-1,n$ provide a different way to compute $\gamma_{2(n-1)},\gamma_{2n}$, which can be shown to be consistent with the results of the previous Subsections.  
\subsection{Critical coupling $g(\epsilon)$}
\label{gofederiv}
In this Subsection we look for the interacting fixed point value of the coupling $g$ at leading order in $\epsilon$. This can be found using the relation $\gamma_{2n-1}=\gamma_1+(n\!-\!1)\epsilon$, only if we knew the anomalous dimension $\gamma_{2n-1}$. The general formula for the anomalous dimension $\gamma_k$ was derived at the beginning of this Section. However, the values $k=2n-2,2n-1$, were excluded there because the correlation function $\braket{\phi\,\phi^{k}\phi^{2n}}$ in these cases would involve a descendent operator, and this questions the use of formula \eqref{3pf1kkp1} which is valid only for primary operators. However, as we will now show by extending the argument used in~\cite{Nii:2016lpa}, at leading order this relation will continue to hold. 
Let us consider $k=2n\!-\!1$. In this case one notices that
\begin{align}
&\braket{\phi(x) \phi (y)^{2n-1} \phi^{2n} (z)  } = \frac{(2n\!-\!1)!}{g  \mu^{(n\!-\!1)\epsilon}}  \Box_y  \braket{\phi(x) \phi (y)\phi^{2n} (z) } \nn \\
&
=\frac{(2n\!-\!1)!}{g  \mu^{(n\!-\!1)\epsilon}}\; \Box_y \frac{C_{1,1,2n}}{|x-y|^{2\Delta_1-\Delta_{2n}} |y-z|^{\Delta_{2n}} |x-z|^{\Delta_{2n}}}\nn\\
&
=\frac{(2n\!-\!1)!}{g  \mu^{(n\!-\!1)\epsilon}}  C_{1,1,2n} \left\{
- \frac{\Delta_{2n} (2\Delta_1\!-\!\Delta_{2n})}{|x-y|^{2\Delta_1-\Delta_{2n}+2} |y-z|^{\Delta_{2n}+2} |x-z|^{\Delta_{2n}-2}} \right. \nn\\ 
& \left. 
+ \frac{(2\Delta_1\!+\!2\!-\!d) \Delta_{2n} }{|x-y|^{2\Delta_1-\Delta_{2n}} |y-z|^{\Delta_{2n}+2} |x-z|^{\Delta_{2n}}} 
+ \frac{(2\Delta_1\!+\!2\!-\!d) (2\Delta_1\!-\!\Delta_{2n})}{|x-y|^{2\Delta_1-\Delta_{2n}+2} |y-z|^{\Delta_{2n}} |x-z|^{\Delta_{2n}}}
\right\} \nn \\  
&
\stackrel{\mathrm{LO}}{=} \frac{(2n\!-\!1)!}{g}  C_{1,1,2n} \,4 \frac{n}{n-1}
\frac{1}{|x-y|^{\Delta_{2n-1}+\Delta_1-\Delta_{2n}} |y-z|^{\Delta_{2n-1}+\Delta_{2n}-\Delta_1} |x-z|^{\Delta_1+\Delta_{2n}-\Delta_{2n-1}}}\,,
\label{3pf12nm21}
\end{align}
since $2\Delta_1\!+\!2\!-\!d=2\gamma_1=O(g^2)$, and $\Delta_{2n\!-\!1}\!-\!\Delta_1=2$. Finally, we insert the leading value of
$C_{1,1,2n}=O(g)$ obtained from Eq.~\eqref{c12k2lm1} with $l=1$ and $k=n$
\be 
C_{1,1,2n} \stackrel{\mathrm{LO}}{=} \frac{g}{(2n\!-\!1)!} \frac{C^{\mathrm{free}}_{1,2n-1,2n} }{n(n-1)(d_{2n}\!-\!2)^2}\,.
\label{c112nm2}
\ee
Recalling that $d_{2n}\!-\!2=2/(n-1)$ we obtain the desired relation, which nevertheless involves a non primary operator,
\begin{equation}
\braket{\phi(x)\phi(y)^{2n-1}\phi^{2n}(z)} \stackrel{\mathrm{LO}}{=}
\frac{C^{\mathrm{free}}_{1,2n-1,2n} }{|x-y|^{\Delta_{2n-1}+\Delta_1-\Delta_{2n}} |y-z|^{\Delta_{2n-1}+\Delta_{2n}-\Delta_1} |x-z|^{\Delta_1+\Delta_{2n}-\Delta_{2n-1}}}\,.
\end{equation}
The same argument can be applied to the other case $k=2n-2$, i.e.\ $\braket{\phi\, \phi^{2n-2} \phi^{2n-1} }$. We can therefore invoke the relation in Eq.~\eqref{special_anomalous}, $\gamma_{2n-1}=\gamma_1+(n\!-\!1)\epsilon$, implied by the constraint on the the scaling dimension of the descendant operator $\phi^{2n-1}$ from the equation of motion, and write
\be
(n\!-\!1)\epsilon +O(g^2)=\gamma_{2n-1}= \frac{2 (n\!-\!1)}{ n!^2} \frac{(2n-1)!}{(n\!-\!1)!}  \Gamma\left(\frac{1}{n-1}\right)^{n-1}\frac{g}{(4 \pi)^n } +O(g^2) \,,
\ee
which gives the linear relation
\be 
g =4 \,c^{1-n} \frac{n!^3}{(2n)!} \epsilon+O(\epsilon^2) =\frac{n!^3}{(2n)!} (4\pi)^n \Gamma \left(\frac{1}{n\!-\!1}\right)^{1-n} \epsilon+O(\epsilon^2) \,.
\ee
It might be interesting to note that using the fixed point value one has access to some features of the theory out of criticality, such as the beta functions. In fact this result is giving the beta function of the dimensionless $g$ for all the multi-critical minimal models at leading order in the $\ep$-expansion.
Taking into account that the leading $g$ at the fixed point is linear in $\ep$, it is possible to uniquely determine
\be 
\beta_g=-(n-1)\ep \,g +  (n\!-\!1) \frac{(2n)!}{n!^3}   \Gamma \left(\frac{1}{n-1}\right)^{n-1}\!\!\!  \frac{g^2}{(4 \pi)^n} +O(g^3)\,. 
\label{gofe}
\ee
which shows that the non trivial fixed point of the CFT is IR attractive ($g>0$).

\subsection{Collecting the results: even potentials}
\label{summary_even}
We summarize the results in this Subsection and give the leading $\epsilon$-dependence of the anomalous dimensions and structure constants found for theories with even potential. 

\subsubsection*{Anomalous dimensions}

The anomalous dimensions $\gamma_k$ for $1\leq k \leq n-1$ are found to be of $O(g^2)$ but only the first two, $\gamma_1$ and $\gamma_2$, are determined at leading order.  The rest are of $O(g)$ and their leading values, together with $\gamma_1$ and $\gamma_2$ are summarized here
\begin{align}
\gamma_1 =2 (n\!-\!1)^2  \frac{n!^6}{ (2n)!^3} \epsilon^2 +O(\epsilon^3)\,,
\label{gamma1e}
\end{align}
\begin{align}
\gamma_2 =8 \frac{(n\!-\!1)^3(n\!+\!1)}{n\!-\!2}  \frac{n!^6}{(2n)!^3} \epsilon^2 +O(\epsilon^3)\,, \quad n>2
\label{gamma2e}
\end{align}
\begin{align}
\gamma_{k}=2 (n\!-\!1)  \frac{n!}{(2n)!} \frac{k!}{(k\!-\!n)!} \epsilon +O(\epsilon^2)\,,  \quad  k \ge n\,.
\label{gammake}
\end{align}
We notice that the expressions \eqref{gamma1e} and \eqref{gammake}, for $\gamma_1$ and $\gamma_k$, are in agreement with the results obtained in \cite{osborn_07} with a perturbative computation.

One may write a generating function for the anomalous dimensions of all these multi-critical theories obtained by $\ep$-expansion around their critical dimensions. 
Such a generator at $O(\ep)$, which gives $\gamma_k$ for any $k\ge n$ in Eq.~\eqref{gammake},
can be written as
\be
F_\gamma^{(\mathrm{even})}(x,y;\ep)
= e^x \left(\sqrt{x y}\sinh{\sqrt{x y}}-2 \cosh{\sqrt{x y}}\right) \ep +O(\ep^2)\,,
\ee
so that one has
\be
\gamma_k(n;\ep)=\left. \frac{\partial^n}{\partial y^n}\frac{\partial^k}{\partial x^k} \right|_{x,y=0}\!\!\! F_\gamma^{(\mathrm{even})} \,.
\ee

\subsubsection*{Structure constants}
It is useful to write the structure constants as a function of $\epsilon$. 
Since they  are defined modulo the normalization of the operator basis we choose to present them
in a scheme where the coefficient of 
the free propagator (at the critical dimension) is  normalized to unity 
and therefore the composite operators are rescaled according to $\phi^k \to \phi^k / \sqrt{c^k} $, where $c$ was defined in Eq.~\eqref{normprop}.

We find
\be 
C_{1,2k,2l-1} =
\frac{n!^3}{(2n)!}\frac{(n\!-\!1)^2}{(k-l)(k-l+1)} \frac{(2k)!(2l\!-\!1)!}{(n\!+\!l\!-\!k\!-\!1)!(k\!+\!n\!-\!l)!(k\!+\!l\!-\!n)!}
\epsilon +O(\epsilon^2)\,,
\label{c1abeven}
\ee
within the limits of Eq.~\eqref{opeset}, namely $k+l\geq n$, $1\!-\!n\!\leq(l-k)\leq n$, $l-k\neq 0, 1$, and
\be \label{c112k_eps}
C_{1,1,2k} =
\frac{(n\!-\!1)^4}{k(k\!-\!1)(k\!-\!n)(k\!-\!n\!+\!1)} \frac{n!^6}{ (2n)!^2}
\frac{(2k)!}{k!^2 (2n\!-\!k\!-\!1)!}
\epsilon^2 +O(\epsilon^3)\,,
\ee
for $k\neq n\!-\!1,n$ and $2\leq k\le 2n-1$. In this scheme all the $\pi$ factors are absent. We note, however, that for comparison with results obtained from perturbation theory other normalizations may prove more convenient.

Let us consider as few explicit examples the cases $n=2,3,4$ which correspond respectively to the {\tt Ising}, {\tt Tricritical} and {\tt Tetracritical} universality classes, and from the set of leading order structure constants that we have found we report all the ones of $O(\epsilon^2)$ and only a few of the infinite sequence of order $O(\epsilon)$. 
For the {\tt Ising} universality class: 
\be 
C_{114} = \frac{2\epsilon}{3}\,, \quad 
C_{125} =\frac{10\epsilon}{3}\,, \quad
C_{136} = 20\epsilon\,, \quad
C_{116} = \frac{5\epsilon^2}{27}\,.
\ee
For the {\tt Tricritical} universality class: 
\be
C_{114} = \frac{3\epsilon}{5}\,, \quad C_{116} = \frac{6\epsilon}{5}\,, \quad C_{125} = 6\epsilon\,, \quad C_{136} = 54\epsilon\,, \nn
\ee
\be
C_{118} = \frac{21\epsilon^2}{5}\,, \quad  C_{1,1,10} = \frac{378\epsilon^2}{125}\,.
\ee
And finally for the {\tt Tetracritical} universality class:
\be 
C_{116} = \frac{18\epsilon}{35}\,, \quad 
C_{118} = \frac{72\epsilon}{35}\,, \quad 
C_{136} = \frac{972\epsilon}{35}\,, \quad
C_{127} = \frac{36\epsilon}{5}\,, \quad
C_{125} = \frac{54\epsilon}{35}\,, 
\ee
\be
 C_{114} =\frac{729\epsilon^2}{6125}\,, \quad
C_{1,1,10} =\frac{26244\epsilon^2}{875}\,, \quad
C_{1,1,12} =\frac{42768\epsilon^2}{875}\,, \quad
C_{1,1,14} =\frac{555984\epsilon^2}{8575}\,. \nn
\ee

\section{\boldmath{$\phi^{2n+1}$}-theory in \boldmath{$d=d_{2n+1}-\epsilon$} dimensions}
\label{section-odd-potentials}

This Section is complementary to Sect.\ \ref{section-even-potentials} in that it is dedicated to the investigation of the odd potentials $\phi^{2n+1}$ for $n$ a natural number $n\geq 1$
which arise as particular cases of \eqref{microscopic_action} by setting $m=2n+1$ for $m$ an odd number.
The odd potentials are not invariant under parity, but are instead protected by a generalization of parity, which has been related to $PT$-symmetry \cite{Bender:2004sa}.
On a general action $S[\phi]$ as in \eqref{microscopic_action_GL} $PT$-symmetry acts as
\begin{equation}
 PT: S[\phi] \to S[-\phi]^\star\,,
\end{equation}
where the star indicates complex conjugation. Invariance under this symmetry implies the $\mathbb{Z}_2$ parity of the previous Section as a special case for all even potentials,
but extends the possible symmetry to incorporate odd potentials, provided that the latter have a purely imaginary critical coupling $g$.
It has been argued that $PT$-symmetry is a valid symmetry, in the sense that it suffices to ensure the stability of the corresponding theory \cite{Bender:2004sa}
and to have a spectrum bounded from below.
On the more pragmatic side, it has been argued that these models interpolate with a well known sequence of minimal \emph{non-unitary} multi-critical models
which begins with the {\tt Lee-Yang} universality class \cite{Cardy:1985yy}.
Starting from the case $n=1$ that corresponds to the {\tt Lee-Yang} class, the upper critical dimensions are
\begin{align}
d_{2n+1} &= 2+\frac{4}{2n-1}= 6 \,,\, \frac{10}{3} \,,\, \frac{14}{5} \,,\, \frac{18}{7} \,,\, \frac{22}{9} \,,\, \dots \,,\, 2\,;
\end{align}
which similarly to $d_{2n}$ tend to two in the limit $n\to\infty$. In a Ginzburg-Landau description these models mark a stark contrast with the even ones:
In fact if the even models can be tuned to criticality by changing their mass,
the odd models must be tuned to criticality by pushing the magnetic field to a critical purely imaginary value \cite{Belavin:2003pu,Koubek:1991dt}.
As a matter of fact these models seem to be non-unitary for all $d\ge 2$.

The well-known upper critical dimension of the {\tt Lee-Yang} universality class is six. All other unversality classes have purely rational upper critical dimensions,
starting from $n=2$, which corresponds to the quintic model $\phi^5$ and which has been named {\tt Blume-Capel} universality class in \cite{Zambelli:2016cbw},
where it has been argued to correspond to a tricritical phase for a Blume-Capel spin system \cite{vonGehlen:1989yn,Mossa:2007fx}.
We want to draw the reader's attention to this latter universality class because its upper critical dimension is bigger than three;
therefore the model provides a less known, but potentially interesting, non-trivial universality class in three dimensions,
and potentially it represents a unique example of a theory that is realized for 
$\epsilon < 1$ 
in a physically interesting scenario.
The models with odd potentials are much less studied than the ones of Sect.\ \ref{section-even-potentials},
thus there will be less room for comparison, 
but we plan to complete their perturbative analysis in a future work \cite{CSVZ2}.
On the other hand, the {\tt Lee-Yang} class is very well known \cite{Fisher:1978pf,Macfarlane:1974vp,deAlcantaraBonfim:1980pe,deAlcantaraBonfim:1981sy,Gracey:2015tta} and we will be able to confirm several CFT quantities in the process.

As for the content of this Section, it will mostly follow the development of Sect.\ \ref{section-even-potentials},
but there will be some important differences.
In the first part we will obtain the explicit leading expressions for the anomalous dimensions $\gamma_1$ and $\gamma_2$ and that $\gamma_k=O(g^2)$.
In the second part we will concentrate on the computation of the structure constants, including $C_{1,1,1}$.
In the third part we will show that the possibility to fix the coupling to its critical value as a function of $\ep$
only occurs for the {\tt Lee-Yang} universality class.
All the results will be summarized in the final part of this Section.

\subsection{Anomalous dimensions } \label{anom_dim_odd}
One can follow exactly the same path of Sect.\ \ref{section-even-potentials} and find the leading relation between $\gamma_1$ and the coupling $g$ by acting with two Laplacians on the propagator and using the SDE, 
which now gives the operatorial relation $\phi^{2n}\sim\Box \phi$  so that $\phi^{2n}$ is a descendant of $\phi$.
Taking into account that the results of Sect.\ \ref{section-even-potentials}  must be shifted as $n \to n+\frac{1}{2}$, so that  $\varphi^{2n} \to \varphi^{2n+1}$, we find
\begin{align}
\gamma_1 =c_{\mathrm{odd}}^{2n-1} \frac{\left(2n\!-\!1\right)^2}{(2n\!+\!1)!} \frac{g^2}{32}+O(g^3)= 
  \frac{(2n\!-\!1)^2}{2 (2n\!+\!1)!} \frac{\Gamma\left(\frac{2}{2n-1}\right)^{2n-1}}{(4\pi)^{2n+1}} g^2+O(g^3)\,,
\label{gamma1_odd}
\end{align}
which for $n=1$ gives the known relation for the {\tt Lee-Yang} universality class \cite{Fisher:1978pf}.
Here $c_\mathrm{odd}$ is obtained from \eqref{normprop} after the shift $n \to n+\frac{1}{2}$,
\be
c_\mathrm{odd}=\frac{1}{4\pi}\frac{\Gamma\left(\frac{2}{2n\!-\!1}\right)}{\pi^{\frac{2}{2n-1}}}\,.
\label{codd}
\ee
Also the derivation of  $\gamma_2$ is straightforward when $n>1$,
since it is based on the form of the correlator $\braket{\phi\,\phi\,\phi^2}$ when all the operators are primary.
Therefore from expression \eqref{gamma2}  we can directly infer
\begin{align}
\gamma_2 =c_{\mathrm{odd}}^{2n-1} \frac{(2n+3)(2n-1)^3}{(2n-3)(2n+1)!}\frac{g^2}{16} +O(g^3)=
 \frac{(2n+3)(2n-1)^3}{(2n-3)(2n+1)!}\; \frac{\Gamma\left(\frac{2}{2n-1}\right)^{2n-1}}{(4\pi)^{2n+1}} g^2+O(g^3)\,,
\label{gamma2odd}
\end{align}
which is valid for $n>1$. Thus  {\tt Lee-Yang} is excluded, but in this case the relation for the first scalar descendant of $\phi$, equation \eqref{special_anomalous} with $m=2n+1$ 
\begin{align}
\gamma_{2n} =\gamma_1+ \frac{2n\!-\! 1}{2}\ep\,,
\label{gamma2ndesc}
\end{align}
comes to rescue and allows the determination of $\gamma_2$ also when $n=1$.

Unfortunately we are not able to find a closed expression for the other anomalous dimensions.
From the study of the correlator  of primary operators
\begin{align}
\braket{\phi(x) \phi^k(y) \phi^{k+1}(z)} = \frac{C_{1,k,k+1}}{|x-y|^{\Delta_{1}\!+\!\Delta_{k}\!-\!\Delta_{k+1}} |y-z|^{\Delta_{k}\!+\!\Delta_{k+1}\!-\!\Delta_{1}} |z-x|^{\Delta_{1}\!+\!\Delta_{k+1}\!-\!\Delta_{k}}}\,,
\label{start3pfoddspecial}
\end{align}
we are now only able  to prove that $\gamma_k=O(g^2)$.
Using the SDE one can relate \eqref{start3pfoddspecial}
to the one which involves the descendant operator $\varphi^{2n}$
\begin{align}
\braket{\phi^{2n}(x) \phi^k(y) \phi^{k+1}(z)} =\frac{(2n)!}{g} \braket{ \Box_x \phi(x) \phi^k(y) \phi^{k+1}(z)}\,.
\label{phi2n}
\end{align}
Acting on \eqref{start3pfoddspecial} with a Laplacian in $x$ and keeping only leading order terms gives
\begin{align}
&\braket{ \Box_x \phi(x) \phi^k(y) \phi^{k+1}(z)} \overset{{\rm  LO}}{=} 2 C^{\rm free}_{1,k,k+1}  \left\{
  \frac{\gamma_1\left(\gamma_1\!+\!\gamma_k\!-\!\gamma_{k+1}\right) }
{ |y-z|^{\frac{4k}{2n-1}} |x-z|^{\frac{4}{2n-1}}} \right.\nn\\ 
&\left.+\frac{4}{2n\!-\!1}
\frac{ \gamma_1}{ |y-z|^{\frac{4k}{2n-1}} |z-x|^{\frac{4}{2n-1}+2}}
   +\frac{2}{2n\!-\!1} \frac{\gamma_1\!+\!\gamma_k\!-\!\gamma_{k+1}} 
{ |x-y|^{2}|y-z|^{\frac{4k}{2n-1}+2} |z-x|^{\frac{4}{2n-1}+2}}
\right\}\,,
\label{3pfgammakodd}
\end{align}
where $C^{\rm free}_{1,k,k+1} = (k+1)! c_{\rm odd}^{k+1} $.
Since the correlator $\braket{ \phi^{2n}  \phi^k \phi^{k+1}}$ is zero in the free theory we can safely assume that it is at least $O(g)$ or smaller.
The bracket terms on the r.h.s of \eqref{3pfgammakodd} are thus $O(g^2)$. Recalling from \eqref{gamma1_odd} that $\gamma_1=O(g^2)$ and considering that perturbative corrections are expressed in terms of integer powers of $g$ we conclude that $\gamma_{k+1}-\gamma_k$ is at least of order $O(g^2)$ and thus 
\be
\gamma_k=O(g^2)\,, \quad\quad k\ge 1\,.
\ee

\subsection{Structure constants}
We will now move on to the analysis of the structure constants. As in the case of even potentials, one can consider the correlation functions $\langle\phi\,\phi^k\phi^l\rangle$ and $\langle\phi\,\phi\,\phi^{2k}\rangle$ with the action of one and two Laplacians respectively. Besides these, in the case of odd potentials the correlation function $\langle\phi\,\phi\,\phi\rangle$ with the action of a triple Laplacian also gives some leading order information on the structure constants. Below, we consider each case in turn.

\subsubsection{Structure constants $C_{1,k,l}$}
In Subsection \ref{C_{1,2k,2l-1}}, for $\mathbb{Z}_2$ symmetric theories, we extracted the possible information 
on $C_{1,k,l}$, $|k-l|\neq 1$ from analysing the related correlation functions. The analysis in the present case for odd potentials goes along the same lines, except that the condition on $k,l$ for the correlator $\braket{\phi^{2n} \phi^k \phi^l}$ to acquire a contribution in the free theory is different. Here $k,l$ have to be either both even or both odd. Furthermore they must satisfy
\be 
\ba{l} k+l-2n\geq 0 \\ l+2n-k\geq 0 \\ k+2n-l \geq 0\,. \ea
\ee
This is equivalent to $k\!+\!l \geq 2n$ and $|l\!-\!k|\le 2n$. In this case the correlator $\braket{\phi^{2n} \phi^k \phi^l}$ in the free theory is
\be
\braket{\phi^{2n}(x) \phi^k(y) \phi^l(z)} \stackrel{\mathrm{LO}}{=} \frac{C^\mathrm{free}_{2n,k,l}}{|x-y|^{(2n+k-l)\delta_{2n+1}} |y-z|^{(k+l-2n)\delta_{2n+1}} |x-z|^{(2n+l-k)\delta_{2n+1}}}\,,
\label{cfree2nkl}
\ee
where 
\be
C^\mathrm{free}_{2n,k,l} =\frac{(2n)! k! l! \, c_\mathrm{odd}^{n+\frac{k+l}{2}}}{\frac{2n+k-l}{2}!\frac{2n+l-k}{2}!\frac{k+l-2n}{2}!}\,.
\ee
Let us therefore consider for $k,l\ne 2n$
\begin{align}
\braket{\phi(x) \phi^k(y) \phi^l(z)} &= \frac{C_{1,k,l}}{|x-y|^{\Delta_{1}\!+\!\Delta_{k}\!-\!\Delta_{l}} |y-z|^{\Delta_{k}\!+\!\Delta_{l}\!-\!\Delta_{1}} |x-z|^{\Delta_{1}\!+\!\Delta_{l}\!-\!\Delta_{k}}}\,.
\label{start3pfodd}
\end{align}
Using the SDE, one can relate this correlator of primary operators, whose form is constrained in a simple way by the conformal symmetry,
to another one which involves a descendant operator and is therefore less simple but can be defined through the relation

\begin{align}
\braket{\phi^{2n}(x) \phi^k(y) \phi^l(z)} =\frac{(2n)!}{g} \braket{ \Box_x \phi(x) \phi^k(y) \phi^l(z)}\,.
\end{align}
This tells that the correlator involving the descendant operator $\phi^{2n}$ gets three contributions with different space-time dependence and three corresponding "structure constants" which depends on the $C_{1,k,l}$, the scaling dimensions $\Delta_{2n}=\Delta_1+2$, $\Delta_k$, $\Delta_l$ and the dimension $d$.

In the following we shall restrict to few considerations based on this relation.
Acting with a Laplacian in $x$ and approximating the exponents in the powers at leading order, one finds
\begin{align}
&\Box_x \braket{ \phi(x) \phi^k(y) \phi^l(z)} = 
    C_{1,k,l}\,\gamma_1 \, \frac{(d\!-\!2)(1\!+\!k\!-\! l)+2(\gamma_1\!+\!\gamma_k\!-\!\gamma_l)}{
|x-y|^{2\frac{2n+k-l}{2n-1}} |y-z|^{2\frac{l+k-1}{2n-1}} |x-z|^{2\frac{l-k+1}{2n-1}}} \nn \\
&   + C_{1,k,l}\, \gamma_1 \frac{(d\!-\!2)(1\!+\!l\!-\! k)+2(\gamma_1\!+\!\gamma_l\!-\!\gamma_k) }{
|x-y|^{2\frac{k-l+1}{2n-1}} |y-z|^{2\frac{l+k-1}{2n-1}} |x-z|^{2\frac{2n+l-k}{2n-1}}} \nn \\
&   +\frac{1}{4} C_{1,k,l}\, \frac{ \left[ (d\!-\!2)(k\!+\!l\!-\! 1)+2(\gamma_k\!+\!\gamma_l\!-\!\gamma_1) \right] \left[ (d\!-\!2)(1\!+\!k\!-\! l)+2(\gamma_1\!+\!\gamma_k\!-\!\gamma_l) \right]}{ 
 |x-y|^{2\frac{2n+k-l}{2n-1}} |y-z|^{-2\frac{2n-k-l}{2n-1}} |x-z|^{2\frac{2n-k+l}{2n-1}} }+\dots 
\label{useful3corr}
\end{align}
One can easily see that the leading contribution comes from the last term, which has indeed the same coordinate dependence of the expression in Eq~\eqref{cfree2nkl}, so that
\be
C_{1,k,l}\stackrel{\mathrm{LO}}{=} \frac{k!\, l! \,\, c_\mathrm{odd}^{n+\frac{k+l}{2}}}{\frac{2n+k-l}{2}!\frac{2n+l-k}{2}!\frac{k+l-2n}{2}!} \frac{(2n\!-\!1)^2}{(k\!-\!l)^2\!-\!1} \frac{g}{4} \,.
\label{c1klodd}
\ee
In particular this is valid for the special case $k=l\geq n$ and gives
\begin{align}
C_{1,k,k} \stackrel{\mathrm{LO}}{=}- \frac{k!^2(2n\!-\!1)^2}{(k\!-\!n)!n!^2} c_\mathrm{odd}^{k+n} \,\frac{g}{4} \,, \qquad k \geq n\,.
\label{c1kkodd}
\end{align}

\subsubsection{Structure constants $C_{1,1,2k}$}
Let us finally consider the correlator $\braket{\phi\, \phi\, \phi^{2k}}$. Again, the analysis in this case follows closely that for the even potentials. Applying box twice to the correlator gives at leading order
\be 
\Box_x\Box_y \braket{\phi(x) \phi(y) \phi^{2k}(z)} \stackrel{\mathrm{LO}}{=}
  \frac{2^6 k(k\!-\!1)(4(k\!-\!n)^2-1)}{(2n\!-\!1)^4} \frac{C_{1,1,2k}}{|x\!-\!y|^{(2n-k)\delta_{2n+1}}  |y\!-\!z|^{k\delta_{2n+1}} |x\!-\!z|^{k\delta_{2n+1}}}\,,
\ee
which has to be compared, as before, with the leading order expression of the correlation function obtained using the SDE twice
\begin{align}
\frac{g^2}{(2n)!^2} \braket{\phi^{2n}(x) \phi^{2n}(y) \phi^{2k}(z)} 
\stackrel{\mathrm{LO}}{=} \frac{g^2}{(2n)!^2}  \frac{C_{2n,2n,2k}^{\mathrm{free}}}{|x-y|^{(2n-k)\delta_{2n+1}}  |y-z|^{k\delta_{2n+1}} |x-z|^{k\delta_{2n+1}}} \,.
\end{align}
This gives the structure constants
\begin{align}
C_{1,1,2k}&\stackrel{\mathrm{LO}}{=}  C_{2n,2n,2k}^{\mathrm{free}} \frac{(2n\!-\!1)^4} {2^6 k(k\!-\!1)(4(k\!-\!n)^2-1)(2n)!^2}  \,g^2\nn\\
&= \frac{(2n\!-\!1)^4} {2^6 k(k\!-\!1)(4(k\!-\!n)^2-1)} \frac{(2k)!}{k!^2 (2n-k)!} c_\mathrm{odd}^{2n+k}\,g^2\,, \qquad 2\le k\le 2n \,.
\label{c112kodd}
\end{align}
Of course the case $k=1$ is excluded from this analysis, and the coefficient $C_{2n,2n,2k}^{\mathrm{free}}$ is nonzero if $k\leq 2n$, therefore the range of validity of this equation is $2\leq k \leq 2n$. For $k=n$ the correlation function under study $\braket{\phi\, \phi\, \phi^{2k}}$ involves a descendent operator and therefore does not have the simple scaling property that we have used above to define $C_{1,1,2k}$. Instead this includes several terms as can be seen by writing
\be 
\braket{\phi(x) \phi(y) \phi^{2n}(z)} = \frac{(2n)!}{g} \Box_z \braket{\phi(x) \phi(y) \phi(z)} = \frac{(2n)!}{g} \Box_z \frac{C_{111}}{|x-y|^{\Delta_1}|y-z|^{\Delta_1}|z-x|^{\Delta_1}}\,.
\ee 
Using Eq.\ \eqref{box3pf1} of Appendix \ref{lapla}, the leading term in this expression can be shown to be 
\be 
-\frac{(2n)!}{g} \frac{\Delta^2_1\, C_{111}}{|x-y|^{\Delta_1-2}|y-z|^{\Delta_1+2}|z-x|^{\Delta_1+2}}\,.
\ee
It turns out that the coefficient of this leading term which we can now call $C_{1,1,2n}$ satisfies Eq.\ \eqref{c112kodd} for $k=n$. This can be seen explicitly by inserting into the above expression the structure constant $C_{111}$ which we compute in the next Subsection. 

\subsubsection{The special case of $C_{111}$ for $n>1$}
Let us now consider the action of a triple Laplacian on $\braket{\phi\, \phi\, \phi}$ for $n> 1$, which lies outside the region of validity of the relation \eqref{c1klodd}. Following the usual argument, by applying the box operator three times one finds the following leading contribution
\be 
\Box_x\Box_y\Box_z \braket{\phi(x) \phi(y) \phi(z)} \stackrel{\mathrm{LO}}{=}\,   \frac{2^8 n(n\!-\!1)}{(2n\!-\!1)^6}  
 \frac{C_{111}}{|x-y|^{\delta_{2n+1}+2}  |y-z|^{\delta_{2n+1}+2} |x-z|^{\delta_{2n+1}+2}}\,,
 \label{box3111}
\ee 
which we can compare with the leading order expression of the same correlator in which the SDE has been used three times
\be 
\frac{g^3}{(2n)!^3} \braket{\phi^{2n}(x) \phi^{2n}(y) \phi^{2n}(z)} 
\stackrel{\mathrm{LO}}{=}\frac{g^3}{(2n)!^3} \frac{C_{2n,2n,2n}^{\mathrm{free}}}{|x-y|^{2n\delta_{2n+1}}  |y-z|^{2n\delta_{2n+1}} |x-z|^{2n\delta_{2n+1}}}\,.
\ee
Comparing the two, we obtain the following expression of order $O(g^3)$ for the structure constant
\begin{align}
C_{111}\stackrel{\mathrm{LO}}{=}\frac{C_{2n,2n,2n}^{\mathrm{free}} } {(2n)!^3} \frac{(2n\!-\!1)^6}{2^8 n(n\!-\!1)}  g^3 =  \frac{c_\mathrm{odd}^{3n} (2n\!-\!1)^6}{2^8 n(n\!-\!1)n!^3}  g^3 \,, \quad n>1\,.
\end{align}

\subsection{Critical coupling $g(\epsilon)$ for $n=1$}
\label{gep_LY_deriv}

If one tries to repeat the argument of the previous Subsection for the case $n=1$, which corresponds to the {\tt Lee-Yang} universality class, the r.h.s of \eqref{box3111} will involve the anomalous dimension $\gamma_1$. Following~\cite{Nii:2016lpa}, one may evaluate at leading order
\be
\Box_x\Box_y\Box_z \braket{\phi(x) \phi(y) \phi(z)} \stackrel{\mathrm{LO}}{=}\,  32(\ep - 6\gamma_1)  \frac{C_{111}}{|x-y|^4 |y-z|^4 |x-z|^4}\,.
\ee
On the other hand, in this case $C_{111}$ is already known, because Eq.~\eqref{c1kkodd} is still valid for $k=1$ and gives\footnote{This is also in agreement with the OPE coefficient found in \cite{Hasegawa:2016piv}} $C_{111}=-c^2_\mathrm{odd} \,g/4=-g/(2\pi)^6$. Therefore comparing this with the corresponding equation found from the SDE
\be 
\frac{g^3}{2!^3} \braket{\phi^2(x) \phi^{2}(y) \phi^{2}(z)} \stackrel{\mathrm{LO}}{=} \frac{g^3}{8} \frac{C_{222}}{|x-y|^4 |y-z|^4 |x-z|^4}\,,
\ee
we find the relation
\be
6\gamma_1-\ep=-\frac{g^3}{8} \frac{C_{222}}{ 32 C_{111}}=\frac{g^2}{8}c_\mathrm{odd} =\frac{g^2}{32\pi^3}\,.
\ee
Recalling from Eq~\eqref{gamma1_odd} that $\gamma_1=(c_\mathrm{odd}/6) g^2/32=g^2/(768\pi^3)$ for $n=1$, one has for the {\tt Lee-Yang} universality class
\be
g^2=-\frac{32}{3\,c_\mathrm{odd}}\ep=-\frac{2}{3}(4\pi)^3\ep\,.
\label{gep_LY}
\ee
Also here this result is giving the beta function of the dimensioneless $g$ for the {\tt Lee-Yang} universality class at  leading order in the $\ep$-expansion.
Taking into account that the leading $g$ at the fixed point is proportional to $\sqrt{-\ep}$, we find that
\be 
\beta_g=-\frac{\ep}{2} g -  \frac{3}{4} \frac{g^3}{(4 \pi)^3} +O(g^4)\,,
\ee
which shows again that the interacting fixed point is IR attractive.

\subsection{Collecting the results: odd potentials}
\label{summary_odd}

Here we collect the various results of this Section. We shall give again the structure constants in the normalization obtained by rescaling the fields $\phi\to\phi  \,c_\mathrm{odd}^{-1/2}$ which normalizes the propagator to unity.

\subsubsection{Case $n=1$, the {\tt Lee-Yang} universality class.}
In the Subsection~\ref{gep_LY_deriv} some relations specific to the {\tt Lee-Yang} ($n=1$ case) have been derived.
Inserting the result for the fixed point of Eq.~\eqref{gep_LY} back into Eq.~\eqref{gamma1_odd}, one finds $\gamma_1$ in terms of $\epsilon$, and finally using the relation \eqref{gamma2ndesc}, which links the anomalous scaling of the descendant operator $\phi^2$ to the one of $\phi$ one obtains the leading $\epsilon$-dependence of $\gamma_2$.
In summary, for the {\tt Lee-Yang} universality class we get
\be
g^2=-\frac{2}{3}(4\pi)^3 \ep\,,\quad \gamma_1=-\frac{1}{18} \ep+O(\ep^2)\,,\quad \gamma_2=\frac{4}{9}\ep+O(\ep^2)\,.
\ee
Moreover, the fact that $\gamma_1+\gamma_k-\gamma_{k+1} = O(g^2)$, shown in Sect.\ \ref{anom_dim_odd}, implies that
\be
\gamma_k=O(\ep) \,.
\ee 
Moving to the structure constants, Eq.~\eqref{c1klodd} for $n=1$ gives  
\be
C_{1,k,l}= \frac{k!\,l!}{\frac{2+k-l}{2}! \frac{2+l-k}{2}! \frac{k+l-2}{2}! } \frac{\sqrt{2/3}}{ (k\!-\!l)^2\!-\!1} \sqrt{-\ep} +O(\ep) \,,\quad |l-k|\leq 2\,.
\ee
In fact one can restrict to $l-k=0,2$, because $k,l$ must be either both even or both odd, so $|k-l|\neq 1$, and the expression is symmetric in $k,l$, so one can take $k<l$ to avoid repetition. Some of these structure constants are listed as follows
\be 
C_{122} =-4\sqrt{\frac{2}{3}}\sqrt{-\epsilon}\,, \quad
C_{111} =-\sqrt{\frac{2}{3}}\sqrt{-\epsilon}\,, \quad  
C_{113}= \sqrt{\frac{2}{3}}\sqrt{-\epsilon}\,, \quad 
C_{133}=-6\sqrt{6}\sqrt{-\epsilon}\,.
\ee
Instead Eq.~\eqref{c112kodd} for $n=1$ gives only one structure constant 
\be
C_{114}=-\frac{\ep}{6}\,. 
\ee

\subsubsection{Case $n>1$}
For the other models, labelled by $n>1$, less information is available from the leading CFT constraints.
It is not possible to find the fixed point $g(\ep)$ so the results are expressed in terms of the coupling $g$,
which always appears through the combination $g \,c_\mathrm{odd}^{n-1/2}$, with $c_\mathrm{odd}$ given in Eq.~\eqref{codd}.

We start from the anomalous dimensions. The leading order constraints give
\begin{align}
\gamma_1&= \frac{\left(2n\!-\!1\right)^2}{(2n\!+\!1)!} \frac{(c_\mathrm{odd}^{n-\frac{1}{2}} g)^2}{32}+O(g^3)\,, \nn\\
\gamma_2 &=\frac{(2n\!+\!3)(2n\!-\!1)^3}{(2n\!-\!3)(2n\!+\!1)!}\frac{(c_\mathrm{odd}^{n-\frac{1}{2}} g)^2}{16}+O(g^3)\,,
\end{align}
from which we can deduce a well determined leading order result for their ratio 
\begin{align}
\frac{\gamma_2}{\gamma_1}
= 2\frac{(2n\!+\!3)(2n\!-\!1)}{(2n\!-\!3)} +O(g)\,. 
\end{align}
While for $k>2$, all one can get is 
\be 
\gamma_k = O(g^2) \,, \qquad  k>2 \,.
\ee
Furthermore, from the relation between the scaling dimension of $\phi$ and $\phi^{2n}$ one finds
\be 
\gamma_{2n} =\gamma_1+ \frac{2n\!-\! 1}{2}\epsilon\,.
\ee
We note that because of the $PT$-symmetry we expect that these models have imaginary fixed point coupling $g(\epsilon)$ 
and therefore we expect both negative $\gamma_1$ and $\gamma_2$ (which is instead positive for the $n=1$ case), at least in the vicinity of the critical dimensions.

For the structure constants we have, at the leading order approximation:
\be
C_{1kl}= \frac{ k! l! }{\frac{2n+k-l}{2}!\frac{2n+l-k}{2}!\frac{k+l-2n}{2}!} \frac{(2n\!-\!1)^2}{(k\!-\!l)^2\!-\!1} 
\frac{ c_\mathrm{odd}^{n-\frac{1}{2}} g}{4} +O(g^2)
\, , \quad k+l\geq 2n\,\, \mathrm{and}\,\, |l-k| \le 2n\,,
\ee
\begin{align}
C_{1,1,2k}&= \frac{(2n\!-\!1)^4} {2^6 k(k\!-\!1)(4(k\!-\!n)^2-1)} \frac{(2k)!}{k!^2 (2n-k)!} ( c_\mathrm{odd}^{n-\frac{1}{2}} g)^2+O(g^3)\,, \quad 2\le k\le 2n \,,
\end{align}
\begin{align}
C_{111}=  \frac{(2n\!-\!1)^6}{2^8 n(n\!-\!1)n!^3}  ( c_\mathrm{odd}^{n-\frac{1}{2}} g)^3 + O(g^4)\,.
\end{align}

\section{Conclusions} \label{conclusions}

We investigated the infinite family of self-interacting scalar theories characterized by a $\phi^m$ potential
using the recent idea proposed by Rychkov and Tan of requiring the compatibility
between conformal invariance and the Schwinger-Dyson equations \cite{Rychkov:2015naa}.
The technique, which was developed further in \cite{Basu:2015gpa,Nii:2016lpa}, allows to express some CFT data as a perturbative expansion in the critical coupling and, for several multi-critical models, also as an $\epsilon$-expansion, where $\epsilon$ is the usual displacement of the dimensionality from its upper critical value $d=d_m-\epsilon$.
What renders our analysis unique is that for most values of $m$, the upper critical dimension is a purely rational number,
making our results more interesting and potentially unexpected.
Our computations agree with the results obtained by O'Dwyer and Osborn through perturbation theory and the renormalization group for $m$ even \cite{osborn_07},
as well as with those obtained in the special cases of {\tt Ising} ($m=4$), {\tt Tricritical} ($m=6$) and {\tt Lee-Yang} ($m=3$)
for which the upper critical dimension is an integer \cite{Rychkov:2015naa, Basu:2015gpa, Raju:2015fza, Nii:2016lpa}.

The sequence of models for $m$ even enjoys $\mathbb{Z}_2$ parity and encodes the scale invariant points for the Ginzburg-Landau description of multi-critical phase-transitions
in which a number $m/2$ of distinct ground states becomes degenerate. These are known to interpolate with the unitary minimal models of CFT in $d=2$.
The sequence of models for $m$ odd enjoys a generalization of parity and is conjectured to interpolate with some non-unitary minimal models in $d=2$ \cite{Zambelli:2016cbw,Belavin:2003pu}.
While there is no formal proof that scale-invariance implies conformal invariance,
we take our results as a pragmatic evidence that conformal invariance could be realized at criticality for the entire sequence of scalar theories that we investigated.
In a future publication we will confirm several results of this paper with an independent computation based on perturbation theory \cite{CSVZ1}.

The extent of our results differs between even and odd models, 
and the strength of the method seems to favour the even potentials.
We dedicated Sect.\ \ref{section-even-potentials} to the even potentials $\phi^{2n}$, for which we could obtain the anomalous dimensions $\gamma_1$ and $\gamma_2$ and $\gamma_{k\ge n}$,
two entire families of structure constant $C_{1,2k,2l-1}$ and  $C_{1,1,2k}$,
as well as a relation between $\epsilon$ and the critical coupling $g(\epsilon)$.
In Sect.\ \ref{section-odd-potentials} we studied the odd potentials $\phi^{2n+1}$,
for which we could determine $\gamma_1$ and $\gamma_2$ together with the structure constants $C_{1,k,l}$, $C_{1,1,2k}$ and $C_{1,1,1}$.
Only for the cubic potential $\phi^3$, corresponding to the {\tt Lee-Yang} universality class, we could find a relation for the critical coupling $g(\epsilon)$.
For all other odd potentials it is however possible to re-express all critical quantities in terms of $\gamma_1$,
which yields some simplification. 
All results are summarized in Sect.s \ref{summary_even} and \ref{summary_odd} 
for even and odd potentials respectively.

Our analysis is very encouraging in that it can be considered as a first step in the perturbative investigation of the CFT data of these unitary and non-unitary multi-critical theories.
In a more general context, the multi-critical models are expected to provide a bridge from criticality in dimension $d\ge 2$ to the well known minimal models in CFT in two dimensions \cite{Codello:2012sc}.
While our results could be compared to the leading results of perturbation theory, the most interesting question that remains open is on how to generalize our use of the CFT constraints
to successfully reproduce higher orders of the $\epsilon$-expansion. It is possible that the correct path is to follow the conformal bootstrap program \cite{ElShowk:2012ht}: possibly using the Mellin space representation \cite{Gopakumar:2016wkt}
and ensuring that the non-unitarity of some theories poses no obstacle \cite{Gliozzi:2014jsa,Pismenskii:2015xxg}, or perhaps exploiting the idea of large spin perturbation theory \cite{Alday:2016njk,Alday:2016jfr}, which may prove useful in this direction.

A special comment must be made on unitarity of the spectrum. In fact, the $\epsilon$-expansion probes the theory for continuous values of the dimensionality, but it has been recently shown
that families of evanescent operators (sometimes associated with total derivatives) appear in the spectrum with negative norms whenever the dimensionality 
is not a natural number.
Furthermore, almost all the $\phi^m$ potentials have a purely rational upper critical dimension.
The role that evanescent operators have on our multi-critical models is still unknown and the presence of negative norm states should be investigated.

The possible non-unitarity of the spectrum should be distinguished from the non-unitarity of the odd potentials, which are characterized by complex values of the coupling constant. 
These odd potentials seem to be protected by a generalization of parity that has been linked to $PT$-symmetry \cite{Bender:2004sa}. 
This manifests in the fact that for all the $n>1$ models one has leading negative $\gamma_1$ and $\gamma_2$ anomalous dimensions
(the latter is positive in the {\tt Lee-Yang} universality class). 
It would be interesting to investigate whether this feature is maintained at higher order in the $\ep$-expansion or at the non pertubative level.
Among all odd models we would like to point out that the quintic model $\phi^5$ has upper critical dimension $d_c=\frac{10}{3}>3$, implying that $\epsilon=\frac{1}{3} < 1$ for $d=3$.
We plan to investigate this model further in the future \cite{CSVZ2}.

\bigskip
\bigskip
\noindent
{\it  Note added:} After the completion of this work we became aware of the two works~\cite{Gliozzi:2016ysv, Gliozzi:2017hni} devoted to the study of generalized Wilson-Fisher critical theories.
One class of models considered there coincides with the multicritical models with even potentials analysed in our Section~\ref{section-even-potentials}. 
With an alternative method based on the expansion of four point correlation functions in conformal blocks, the Authors were able to provide some of the results found here.
In particular the leading anomalous dimensions $\gamma_1$ and $\gamma_k$ for $k> n$.
Moreover they find (the square of) a family of leading OPE coefficients (see Eq. (4.36) of~\cite{Gliozzi:2017hni}) which coincides with our Eq.~\eqref{c1abeven},
once the composite operators $\phi^k$ are rescaled by $\sqrt{k!}$ in order to have their two point correlation function normalized to unity.
In Section~\ref{section-even-potentials}, in addition to these overlapping material, which are however obtained by different approaches, we have provided the leading value of $\gamma_2$ for $n>2$, given in our Eq.~\eqref{gamma2e}, as well as the independent family of $\mathcal{O}(\epsilon^2)$ structure constants $C_{1,1,2k}$ that we have reported in Eq.~\eqref{c112k_eps}.


\smallskip
\bigskip

\noindent
{\bf Acknowledgments}\\
We are grateful to M.\ Ammon for sharing useful comments on CFT and reading an earlier version of this draft.
O.Z.\ acknowledges support by the DFG under grant No.~Gi328/7-1.
A.C.\ and O.Z.\ are grateful to INFN Bologna for hospitality and support.

\appendix
\numberwithin{equation}{section}
\section{Free theory}
\label{free}
Important inputs that have been used in the calculation are the expression of two and three point correlators for a free theory,
which are usually computed using the Wick theorem (Gaussian path integrals).
We give here some general relations that are used in the text.

The propagator of the free theory at the critical dimension $d_m=2(1\!+\!\delta_m)$ is given by
\begin{align}
\braket{\phi(x)  \phi(y)} \overset{\mathrm{free}}=\frac{c}{|x-y|^{2 \delta_m}}\,,
\end{align}
where 
\begin{align}
c=\frac{1}{4 \pi} \frac{\Gamma(\delta_m)}{\pi^{\delta_m}}=\frac{1}{(d_m\!-\!2)S_{d_m}} \,.
\label{normprop}
\end{align}
Here $S_{d_m}$ is the area of the $d_m$-dimensional sphere.
A generic two point correlator for the operators $\phi^k$ is given by
\begin{align}
\braket{\phi^k(x)  \phi^l(y)} \overset{\mathrm{free}}{=}\delta_{k l} \,  k!  \frac{c^k}{|x-y|^{2 k \delta_m}}\,,
\label{2pf}
\end{align}
where the $k!$ counts the numbers of possible contractions. As commonly done for a CFT one can rescale the fields to obtain two point functions normalized to one.

\begin{figure}
\begin{center}
\includegraphics[width=5cm]{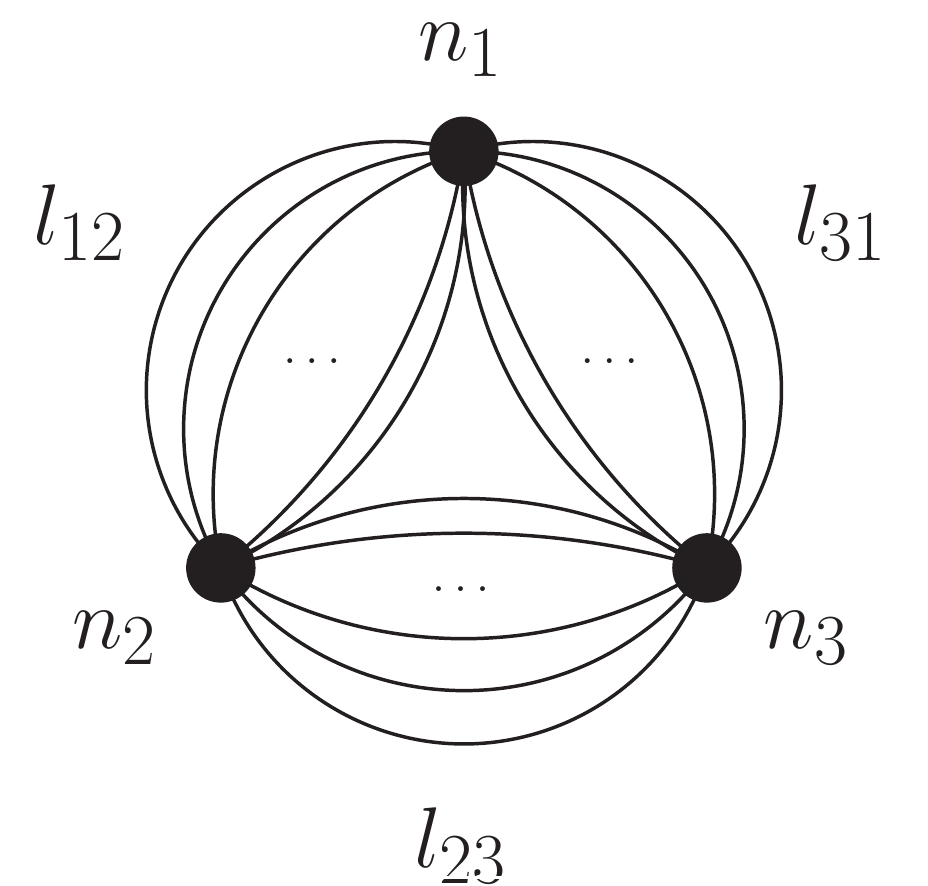}
\end{center}
\caption{Wick contraction counting of a three point correlator.
The vertices are labelled by $i=1,2,3$, the order of the $i$-th vertex is $n_i$, and there are $l_{ij}$ lines connecting two distinct vertices $i$ and $j$.}
\label{3pfcounting}
\end{figure}

We finally consider a generic three point correlator of the form
\begin{align}
\braket{\phi^{n_1}(x_1)  \phi^{n_2}(x_2) \phi^{n_3}(x_3)} \,.
\end{align}
The first constraint for a non zero correlator is that $(n_1+n_2+n_3)\!\! \mod 2=0$, i.e.\ the sum of the powers must be even.
The explicit form of the tree level correlator can be written easily. One can visualise it as a three point diagram (see Fig.~\ref{3pfcounting}) 
with vertices of order $n_1$, $n_2$ and $n_3$ connected by $l_{12}$, $l_{23}$ and $l_{31}$ propagators, in cyclic order respectively.  One has three constraints relating the $n_k$ and the $l_{ij}$ for $i \ne j \ne k$:
\be
n_i=l_{ij}+l_{ki}  \Longleftrightarrow l_{ij} = \frac{1}{2}\left(n_i+n_j-n_k\right)   \, , \qquad i \ne j \ne k\,.
\ee
The correlator is non zero when there exists a solution such that $l_{ij}$ are non negative integers ($l_{ij}\ge 0$).
Then the number of all possible configurations (contractions) is given by the possible splittings (combinations) of $n_i$ in pairs $l_{ij}$ and $l_{ki}$, for each vertex, multiplied by the
possible permutations within each group $l_{ij}$ of contractions. This leads to the counting
\be
N_{n_1,n_2,n_3}= \frac{n_1! \ n_2! \ n_3!}{l_{12}! \ l_{23}!\  l_{31}!}
\ee
so that, with the above normalization, the explicit form of the correlator is given by
\begin{align}
\hspace{-0.5cm} \braket{\phi^{n_1}(x_1)  \phi^{n_2}(x_2) \phi^{n_3}(x_3)}& \overset{\mathrm{free}}{=} 
  \frac{C_{n_1,n_2,n_3}^{\mathrm{free}}}{|x_1\!-\!x_2|^{\delta_m (n_1+n_2-n_3)}|x_2\!-\!x_3|^{\delta_m (n_2+n_3-n_1)}
  |x_3\!-\!x_1|^{\delta_m (n_3+n_1-n_2)}} \,,
\label{3pf_free}
\end{align}
where
\begin{align}
C_{n_1,n_2,n_3}^{\mathrm{free}} &= 
 \frac{n_1! \ n_2! \ n_3!}{\left(\frac{n_1+n_2-n_3}{2}\right)!\left(\frac{n_2+n_3-n_1}{2}\right)!\left(\frac{n_3+n_1-n_2}{2}\right)!} c^{\frac{n_1+n_2+n_3}{2}} \,.
\label{c3_free}
\end{align}

\section{Action of the Laplacian}
\label{lapla}

We give here few useful formulae for the action of one and two Laplacians on two and three point correlators
which are used several times in the computations. 
Starting from the simple relations $\partial_{x^\mu} |x\!-\!y|^{-\alpha}=-\alpha(x\!-\!y)_\mu  |x\!-\!y|^{-\alpha-2}$ and
 \begin{align}
 \Box_x \frac{1}{|x\!-\!y|^{\alpha}}= \frac{\alpha (\alpha\!+\!2\!-\!d)}{|x\!-\!y|^{2+\alpha}}\,,
 \end{align}
one first derives 
 \begin{align}
 \Box_x \Box_y \frac{1}{|x-y|^{\alpha}}
= \frac{\alpha(\alpha \!+\!2) (\alpha\! +2-\!d)  (\alpha\!+\!4\!-\!d)}{|x-y|^{\alpha+4}}   \,.
\label{2pfbbappendix}
\end{align}
We can directly apply the above relations to the coordinate dependent form of three point correlators and find some lengthy expressions.
The action of one Laplacian $\Box_x$ is:
\begin{align}
\Box_x & \frac{1}{|x-y|^{\alpha_1} |y-z|^{\alpha_2} |x-z|^{\alpha_3}}= 
\frac{\alpha_1(\alpha_1\!+\!\alpha_3\!+\!2\!-\!d)}{|x-y|^{\alpha_1+2} |y-z|^{\alpha_2} |x-z|^{\alpha_3}}  \nonumber \\
&+  \frac{\alpha_3(\alpha_1\!+\!\alpha_3\!+\!2\!-\!d)}{|x-y|^{\alpha_1} |y-z|^{\alpha_2} |x-z|^{\alpha_3+2}}
- \frac{\alpha_1 \alpha_3}{|x-y|^{\alpha_1+2} |y-z|^{\alpha_2-2} |x-z|^{\alpha_3+2}}\,.
\label{box3pf1}
\end{align}
The action of two Laplacians $\Box_x \Box_y$ is:
\begin{align}
\Box_x \Box_y & \frac{1}{|x-y|^{\alpha_1} |y-z|^{\alpha_2} |z-x|^{\alpha_3}} = 
\frac{\alpha_2 \alpha_3 (\alpha_1\!+\!\alpha_2\!+\!2\!-\!d) (\alpha_1\!+\!\alpha_3\!+\!2\!-\!d)}{|x-y|^{\alpha_1} |y-z|^{\alpha_2+2} |x-z|^{\alpha_3+2}}\nonumber\\
&+\frac{\alpha_1 \alpha_2 (\alpha_1\!+\!\alpha_3\!+\!2\!-\!d)(\alpha_1\!+\!\alpha_2\!-\!\alpha_3\!+\!4\!-\!d)}{|x-y|^{\alpha_1+2} |y-z|^{\alpha_2+2} |x-z|^{\alpha_3}} 
+\frac{\alpha_1 \alpha_3 (\alpha_1\!+\!\alpha_2\!+\!2\!-\!d)(\alpha_1\!+\!\alpha_3\!-\!\alpha_2\!+\!4\!-\!d)}{|x-y|^{\alpha_1+2} |y-z|^{\alpha_2} |x-z|^{\alpha_3+2}} \nonumber\\
&-\frac{\alpha_1  \alpha_2 (2+\alpha_1) (\alpha_1\!+\!\alpha_3\!+\!2\!-\!d)}{|x-y|^{\alpha_1+4} |y-z|^{\alpha_2+2} |x-z|^{\alpha_3-2}} 
-\frac{\alpha_1  \alpha_3 (2+\alpha_1) (\alpha_1\!+\!\alpha_2\!+\!2\!-\!d)}{|x-y|^{\alpha_1+4} |y-z|^{\alpha_2-2} |x-z|^{\alpha_3+2}} 
 \nonumber \\
& + \frac{\alpha_1 (2+\alpha_1)\left(2 \alpha_2 \alpha_3 +(\alpha_1\!+\!2\!-\!d)(\alpha_1\!+\!\alpha_2\!+\! \alpha_3+\!4\!-\!d)\right)}{|x-y|^{\alpha_1+4} |y-z|^{\alpha_2} |x-z|^{\alpha_3}} \,.
\label{boxbox3pf11}
\end{align}
A similar but much longer expression can be derived applying a third Laplacian to the three point function at $z$
and although it is used in Sections \ref{section-even-potentials} and \ref{section-odd-potentials} we will not report it here.

\end{document}